%% file: main.tex
\documentclass[fleqn,usenatbib]{mnras}

\usepackage{newtxtext,newtxmath}

\usepackage[T1]{fontenc}

\DeclareRobustCommand{\VAN}[3]{#2}
\let\VANthebibliography\thebibliography
\def\thebibliography{\DeclareRobustCommand{\VAN}[3]{##3}\VANthebibliography}

\usepackage{graphicx}	
\usepackage{amsmath}	

\title[A knotted and slower Milky Way bar]{Disentangling the Galaxy's Gordian knot: evidence from \textsl{APOGEE-Gaia} for a knotted and slower bar in the Milky Way}

\author[D. Horta et al.]{
Danny Horta$^{1,2}$\thanks{E-mail: dhortadarrington@flatironinstitute.org}, 
Michael S. Petersen$^{2}$,
Jorge Pe\~{n}arrubia$^{2}$
\\
$^{1}$Center for Computational Astrophysics, Flatiron Institute, 162 5th Ave., New York, NY 10010, USA\\
$^{2}$Institute for Astronomy, University of Edinburgh, Royal Observatory, Blackford Hill, Edinburgh, EH9 3HJ, UK\\
}

\date{Accepted XXX. Received YYY; in original form ZZZ}

\pubyear{2024}

\input{preamble.tex}

\begin{document}
\label{firstpage}
\pagerange{\pageref{firstpage}--\pageref{lastpage}}
\maketitle

\begin{abstract}

The inner $\sim5$ kiloparsec (kpc) region of the Milky Way is complex. Unravelling the evolution of the Galaxy requires precise understanding of the formation of this region. We report a study focused on disentangling the inner Galaxy ($r<5~\kpc$) using the measured positions, velocities, and element abundance ratios of red giant stars from the \textsl{APOGEE-Gaia} surveys.  After removing the stellar halo, inner Galaxy populations can be grouped into three main components based on their angular momentum: bar, disc, and a previously unreported ``knot'' component. The knot has a spheroidal shape, is concentrated in the inner $\sim1.5~\kpc$, is comprised of stars on nearly-radial orbits, and contains stars with super-solar [Fe/H] element abundances. The chemical compositions of the knot are qualitatively similar to the Galactic bar and inner disc, suggestive that these three populations share a common genesis; the chemical/dynamic properties of the knot suggest it could constitute a classical bulge formed via secular evolution. Moreover, our results show that the bar is more slowly rotating than previously thought, with a pattern speed of $\Omega_{\mathrm{bar}}=24\pm3~\kms~\kpc^{-1}$. This new estimate suggests that the influence of the bar extends beyond the solar radius, with $R_{\mathrm{CR}}\sim9.4-9.8~\kpc$, depending on the adopted Milky Way rotation curve; it also suggests a ratio of corotation to bar length of $\mathcal{R}\sim1.8-2$. Our findings help place constraints on the formation and evolution of inner Galaxy populations, and directly constrain dynamical studies of the Milky Way bar and stars in the solar neighbourhood.

\end{abstract}




\section{Introduction}

The inner regions of galaxies, commonly referred to as ``bulges'', are complex. This is because they are comprised by the superposition of many Galactic components (e.g., disc, stellar halo, bar). However, these inner regions are key to understand to fully unravel the formation and evolution of galaxies (\citealp{Wyse1997,Prugniel2001,Barbuy2018}). 

For the case of \textit{the Galaxy} (the Milky Way), early balloon observations of near-infrared emission \citep{Blitz1991} and studies of the kinematics of gas \citep{Peters1975,Binney1991} in the inner Galactic neighbourhoods revealed that the Milky Way hosts a coherently rotating bar. However, the vast amount of interstellar extinction obscuring this region has made direct studies of the inner Galaxy difficult. As a work-around, studies have focused on instead measuring the properties of stars in fields typically away from the Galactic mid-plane, where extinction is less prominent. These have confirmed that the inner few kiloparsecs are dominated by a coherently rotating population (i.e., a bar) and a boxy/peanut/X-shaped distribution (\citealp[e.g.,][]{Lopez2005,Rattenbury2007,Nataf2010,McWilliam2010,Saito2011,Ness2013,Ness2013b,Wegg2013,Portail2015}) conjectured to possibly be the result of a `buckling' bar \citep[e.g.,][]{Debattista2004,Kormendy2004,Athanassoula2005}; this leaves little to no room for a ``classical'' bulge component \footnote{Defining a classical bulge component has been an exercise that dates back almost seven decades \citep{Wyse1997}. However, typically, a classical bulge is defined as an old and metal-rich (super-solar metallicities) component, that is usually preconceived to be related to elliptical galaxies \citep{Davies1983,Whitford1986,Wyse1997}.}. 

For this reason, measuring the properties of the Milky Way's bar, such as the pattern speed (\citealp[][]{Dehnen1999,Wang2013,Portail2017,Sanders2019,Bovy2019,Binney2020}), length \citep[][]{Wegg2015,Lucey2023}, and its 3D structure \citep[e.g.,][]{Saito2011,Wegg2013,Wegg2015} have been attempted over and over in the literature, as its been shown that these measurables strongly influence the observed perturbed kinematics of stars in the disc (\citealp[e.g.,][]{Dehnen2000,Minchev2010,Hunt2018,Trick.2019,Fragkoudi2019,Fijii2019}) and even on the structure of nearby stellar streams \citep[][]{Pearson2017, Thomas.2023} and the stellar halo \citep{Dillamore.2023}.

Moreover, in addition to the dynamical and structural properties of the Galactic bar, several attempts have been made to infer its chemical composition thanks to data supplied by large spectroscopic surveys. This has enabled the chemical characterisation of inner Galaxy populations, showing that the bar is chemically similar to its co-spatial disc counterpart, following a trajectory that spans a high [$\alpha$/Fe] loci at sub-solar [Fe/H], and reaches a low [$\alpha$/Fe] abundance at super-solar [Fe/H] \citep{Barbuy2018,Bovy2019,Queiroz2020,Debattista2023}. Such chemical similarity naturally implies that the bar must be related to the disc of the Milky Way and formed over a similar timescale to the inner disc. Moreover, all inner Galaxy stellar populations appear to be comprised primarily by old stars \citep[e.g.,][]{Hasselquist2020}, although a tail of young stars has also been detected \citep[e.g.,][]{Bensby2017}.

In this paper, we present an application of a method \citep{Penarrubia2021} that is able to statistically decompose stellar populations within $\sim5~\kpc$ of the Galactic Centre using angular momentum information. We apply this method to the latest \textsl{APOGEE} (DR17) and \textit{Gaia} (DR3) data (Section~\ref{sec_data}), and fit a three component Gaussian Mixture Model (GMM) to the angular momentum distribution of disc stars within $\sim5~\kpc$ from the Galactic Centre (Section~\ref{sec_method}). From these fits, we find that the data can be statistically divided into three separate (dynamic) components: a ``knot'', a bar, and an inner disc. We then characterise the chemical properties of the components identified to infer their origin, and empirically measure the pattern speed of the Milky Way bar in Section~\ref{sec_results}. We discuss our findings within the context of previous work in Section~\ref{sec_discussion} and provide our conclusions in Section~\ref{sec_conclusions}. To ascertain the robustness of this methodology, we also test our procedure on mock catalogues created from an N-body simulation of a Milky Way-mass galaxy with a bar in Appendix~\ref{sec_mock}, finding quantitatively similar results to the observational data. These checks ascertain the robustness of our method.

\section{Data}
\label{sec_data}
This paper uses a cross-matched catalogue of the latest data release \citep[DR17,][]{sdss2021} of the SDSS-III/IV (\citealp{Eisenstein2011,Blanton2017}) and \textsl{APOGEE} survey (\citealp[][]{Majewski2017}) with distances and astrometry determined from \textit{Gaia} \citep[DR3,][]{Gaia2020, Bailer2021}. The radial velocities supplied by \textsl{APOGEE} \citep[][Holtzman et al, in prep]{Nidever2015}, when combined with the sky positions, proper motions, and inferred distances \citep{Bailer2021} based on \textit{Gaia}\footnote{We note that we also tried using other sets of distances (e.g., \textsl{astroNN}: \citealp[][]{Leung2019b}; \textsl{StarHorse}: \citealp[]{Anders2022}) and found that our results remain unchanged.}, provide complete 6-D phase space information for $\sim$650,000 stars in the Milky Way, for most of which exquisite elemental abundance ratios for up to $\sim$20 different elemental species have been determined. All \textsl{APOGEE} data are based on observations collected by (almost) twin high-resolution multi-fibre spectrographs \citep{Wilson2019} attached to the 2.5m Sloan telescope at Apache Point Observatory \citep{Gunn2006}
and the du Pont 2.5~m telescope at Las Campanas Observatory
\citep{BowenVaughan1973}. Elemental abundances are derived from
automatic analysis of stellar spectra using the \textsl{ASPCAP}
pipeline \citep{Perez2015} based on the \textsl{FERRE}\footnote{github.com/callendeprieto/ferre} code \citep[][]{Prieto2006} and the line lists from \citet{Cunha2017} and \citet[][]{Smith2021}. The spectra themselves were
reduced by a customized pipeline \citep{Nidever2015}. For details on
target selection criteria, see \citet{Zasowski2013} for \textsl{APOGEE}, \citet{Zasowski2017} for \textsl{APOGEE}-2, \citet[][]{Beaton2021} for \textsl{APOGEE} north, and \citet[][]{Santana2021} for \textsl{APOGEE} south. \textit{Gaia} data supplies sky positions, proper motion, and parallax measurements for $\sim 2$ billion stars.

To convert between astrometric parameters and cartesian coordinates, we use the 6-D phase space information\footnote{The positions, proper motions, and distances are taken/derived from \textit{Gaia} DR3 data, whilst the radial velocities are taken from \textsl{APOGEE} DR17.} assuming a solar velocity combining the proper motion from Sgr~A$^{*}$ \citep{Reid2020} with the determination of the local standard of rest of \citet{Schonrich2010}. This adjustment leads to a 3D velocity of the Sun equal to [U$_{\odot}$, V$_{\odot}$, W$_{\odot}$] = [--11.1, 248.0, 8.5] km s$^{-1}$. We assume the distance between the Sun and the Galactic Centre to be R$_{0} = 8.178~\kpc$ \citep{Gravity2019}, and the vertical height of the Sun above the midplane $z_{0} = 0.02~\kpc$ \citep{Bennett2019}.

\subsection{Parent sample}
The parent sample used in this work comprised of stars in \textsl{APOGEE}-\textit{Gaia} that pass the following criteria:
\begin{itemize}
    \item \textbf{High signal-to-noise:} S/N$>50$.
    \item \textbf{Red giant stars with good \textsl{\textsl{APOGEE}} stellar surface parameters:} $3500<T_{\mathrm{eff}}<5500$ and $3>\logg>0$.
    \item \textbf{High-quality stellar parameters:} uncertainties in the surface gravity below $\logg_{\mathrm{err}}<0.2$ and \textsl{APOGEE} \texttt{STARFLAG} bits \textit{not} set to 0, 1, 3, 16, 17, 19, 21, 22, \textsl{APOGEE} \texttt{ASPCAPFLAG} bits \textit{not} set to 23, and \texttt{EXTRATARG} equal to 0.
    \item \textbf{Reliable distance measurements:} \citet{Bailer2021} distances with $d_{\odot}$/$d_{\odot,\mathrm{err}}$ $>5$ (i.e., $\leq 20\%$ uncertainty).
    \item \textbf{No halo stars:} $\mathrm{[Fe/H]}>-0.8$.
    \item \textbf{Stars in the inner Galaxy:} $r<5~\kpc$.
    \item \textbf{No globular cluster stars:} Remove all members using the catalogues from \citet[]{Horta2020} and \citet[][]{Schiavon2023}.
\end{itemize}

\begin{figure*}
    \centering
    \includegraphics[width=\textwidth]{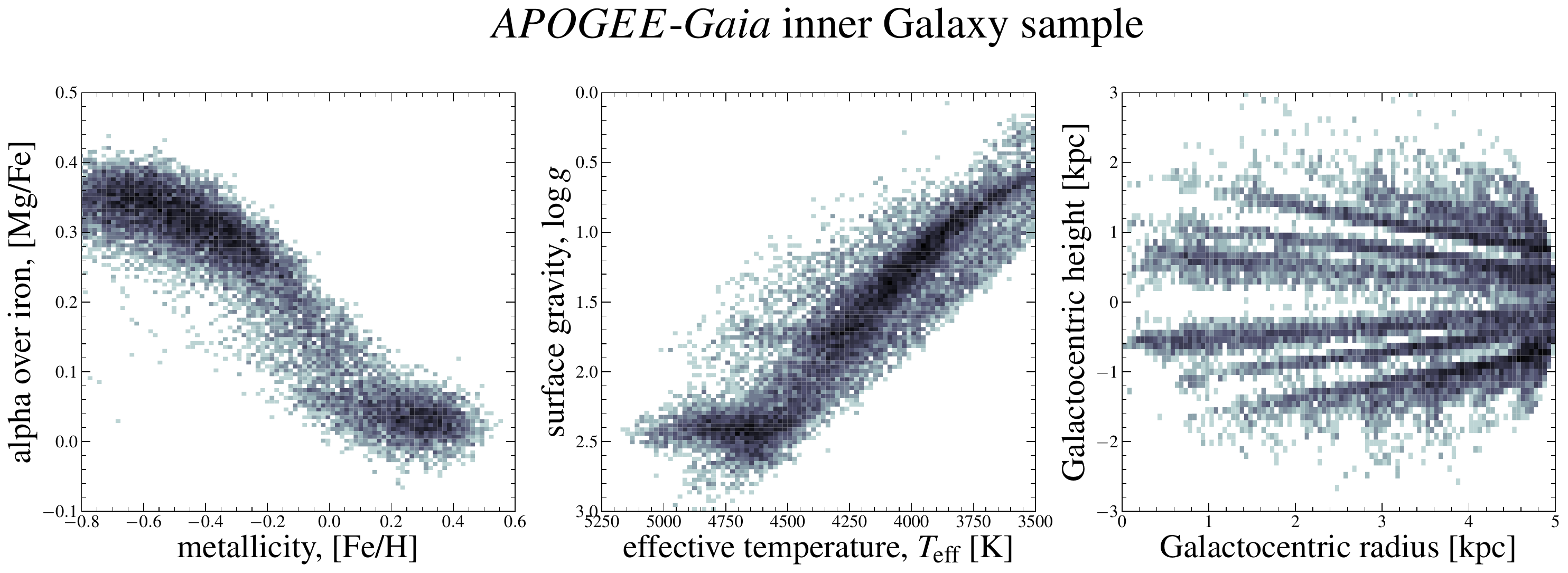}
    \caption{ 2D density distribution of the parent sample in the [$\alpha$/Fe]-metallicity plane (\textit{left}), the surface gravity vs effective temperature diagram (\textit{middle}), and the Galactocentric cylindrical positions (\textit{right}).}  \label{fig_data}
\end{figure*}

Our selection criteria leads to 18,933 stars with reliable elemental abundance rations, distances, positions, and velocities. In the following Section, we will use this sample and decompose it in angular momentum space in an attempt to disentangle metal-rich stellar populations in the inner Galaxy. 

\section{Method}
\label{sec_method}

\subsection{Angular momentum decomposition with Gaussian mixture likelihoods}

Before describing in full the procedure employed for decomposing the inner Milky Way with Gaussian mixture likelihoods, we summarise the main steps in our methodology:

\begin{enumerate}
\item We first determine our parent sample of inner Galaxy stars using the cross-match from \textsl{APOGEE} and \textsl{Gaia} and the flags described in Section~\ref{sec_data}.
\item We define our likelihood using a Gaussian Mixture Model (GMM) as described below and in detail in \cite{Penarrubia2021}. Briefly, the GMM approximates the angular momentum (\textbf{L}) distribution of a sample of stars using a number of Gaussians ($K$).
\item To optimise the model parameters, we maximise the likelihood function (or minimise the log-likelihood) using the \texttt{multinest} sampling technique \citep{multinest}. This step requires setting priors for the model parameters. 
\item Using the optimised model parameters, we compute the probability of a point in angular momentum space (i.e. an observed star) being associated with a given component based on its \textbf{L} vector and the optimised model parameters.  
\end{enumerate}

In Appendix~\ref{sec_mock}, we validate our method by constructing a mock \textsl{\textsl{APOGEE}} catalogue using a simulation of a realistic barred Milky Way-mass galaxy \citep{Petersen.Weinberg.Katz.2021} and performing our GMM decomposition. We find that the method recovers the distinct dynamical components and their properties to a high degree of accuracy.

In more detail, our method begins with the creation of a constrained Gaussian mixture model to approximate the distribution of stars in angular momentum (\textbf{L}) space. We construct a likelihood for which we maximise the evidence  given the data (consisting of $N$ stars) and model (consisting of $K$ multivariate Gaussians). The method is almost identical to that of \cite{Penarrubia2021}, with two key modifications: here, we allow for correlations between two angular momentum dimensions, and also `sort' the angular momentum space by defining the mean angular momentum modulus of each component relative to the component with the largest mean angular momentum. We describe these changes to the log-likelihood, and summarise how to compute membership probabilities for any star given the fitted model.

We start by defining the indices $\kappa\in[1,K]$ to specify the number of components. We choose $K=3$ as the number of components that best describes the data, after trialling for $K=[1,2,3,4]$. As in \cite{Penarrubia2021}, the data is described by the vector 
\begin{equation}
    \vec{\theta}=\{\ell,b,D,\mu_{\ell},\mu_{b},v_{{\rm 
los}},\epsilon_{D},\epsilon_{\mu_{\ell}},\epsilon_{\mu_{b}},\epsilon_{v_{{\rm 
los}}},\rho_{\epsilon_{\mu_{\ell}}\epsilon_{\mu_{b}}}\}_{\nu=1}^{N_{\rm sample}}.
\end{equation}

\begin{figure*}
    \centering
    \includegraphics[width=1\textwidth]{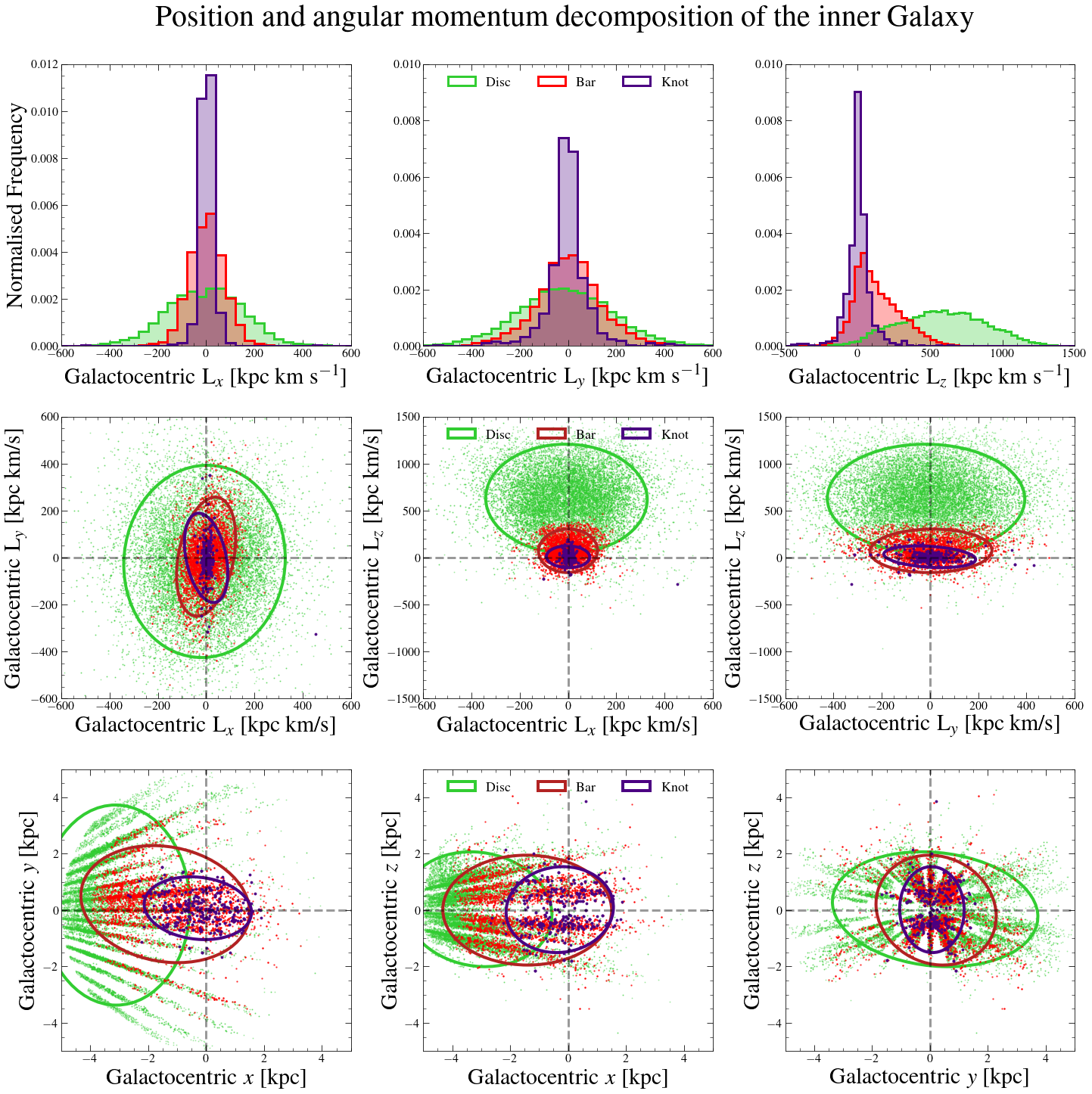}
    \caption{\textit{Top row}: Normalised angular momentum distribution of the parent sample weighted by the probability of stars being associated with the knot (blue), bar (red), and disc (green). \textit{Middle row}: Angular momentum distribution of the three samples identified using our angular momentum decomposition Gaussian mixture likelihood method with their corresponding 2$\sigma$ confidence ellipse. Here, the confidence ellipse should not be interpreted as the Gaussian fits to the data, but are instead to guide the eye as to the extent of each component.
    The distribution of these components is much tighter in angular momentum space than in position space, making it easier to dissect stellar populations in the inner Galaxy. For these and the bottom row, only stars with high probability of them being associated with their respective components are shown (i.e., $p>0.8$). The knot component shows a much tighter distribution in angular momentum, centred around $\boldmatrix{L}\sim\!0~\kpc~\kms$. \textit{Bottom row}: Galactocentric Cartesian coordinate projections of the three components (disc, bar, and knot) identified with the angular momentum decomposition with their corresponding 2$\sigma$ confidence ellipses. The \textsl{\textsl{APOGEE}} pencil-beam footprint is clearly apparent, making any identification in [$x,y,z$] space challenging. In the middle and bottom rows the cross-hair corresponds to the position of the Galactic centre.}  \label{summary}
\end{figure*}

Uncertainties are denoted by $\epsilon$, and correlations between uncertainties (of which we only retain the correlation between the proper motion observations) are denoted by $\rho$. We assume negligible uncertainty on the positions $(\ell,b)$. To separate the Gaussians corresponding to each component with their mean angular momentum value, we perform the fit using sorting parameters $\{\log_{10}\langle L_1 \rangle,\xi_2,\xi_3\}$ to ensure that the components are distinct in total angular momentum. Explicitly, $\log_{10} \langle L_1\rangle$ is the logarithm of the modulus of the mean angular momentum vector of the highest angular momentum component (i.e., total angular momentum $\langle L_1 \rangle$), $\xi_2$ is the prefactor that scales the total angular momentum of $\langle L_1 \rangle$ to calculate the total angular momentum of the second component ($\langle L_2\rangle=\xi_2\langle L_1 \rangle$), and $\xi_3$ is the prefactor that scales the total angular momentum of the second component to calculate the total angular momentum of the third component (($\langle L_3\rangle=\xi_3\langle L_2\rangle=\xi_2\xi_3\langle L_1\rangle$). This way, the fitted components always have stratified angular momenta, and confusion between components is reduced. In addition to the mean angular momentum, each of the populations has parameters to describe the amplitude, location, width, and $x-y$ correlation ($\gamma$, see below) of the Gaussian, $\{f^\kappa,\phi^\kappa,\cos \theta^\kappa,\sigma_{L_x}^\kappa,\sigma_{L_y}^\kappa,\sigma_{L_z}^\kappa,\gamma^\kappa\}_{\kappa=1}^{3}$.
This results in a total of 24 parameters for the mixture model. Here, $f^\kappa$ is the fraction of stars in the sample in the $\kappa$-th component and $(\phi^\kappa,\theta^\kappa)$ are the angles between the angular momentum vector components in spherical coordinates.
$(\sigma^\kappa_{L_x},\sigma^\kappa_{L_y},\sigma^\kappa_{L_z}$) are hyperparameters fitted in inverse squared space for ease of numerical computation and added in quadrature to the observational uncertainties. Thus, the quadrature sum of these uncertainty hyperparameters, $\sqrt{\sigma_{L_{x,y,z}}}$, provide an estimate of the intrinsic  spread of each Gaussian (i.e. after observational uncertainties have been accounted for).
Finally, since there is no a priori guarantee that the symmetry axes of the $\kappa$-th component are aligned with the axis joining the Sun to the Milky Way centre, we introduce the rotational angles $\gamma^\kappa$ around the $z-$axis as additional hyperparameters in our analysis. We adopt flat priors on $\{f,\log\langle L_1\rangle,\xi_2,\xi_3,\phi,\cos\theta,\gamma\}$, and Jeffreys priors for the Gaussian widths $\sigma$, with ranges that include reasonable values: $f\in[0,1]$, $\log\langle L_1\rangle\in[1,4.5]$, $\xi_{2,3}\in[0,1]$, $\phi\in[-\pi,\pi]$, $\cos\theta\in[-1,1]$, $\gamma\in[0,\pi/2]$, and $1/\sigma^2\in[10^{-8},10^{-2}]$.

For each star, we compute the uncertainties in the angular momentum components via a Monte Carlo step\footnote{We set $M=1000$ as the number of draws. We have verified that $\boldmatrix{L}$ is converged with our choice of $M$. With $\textbf{L}_x^\nu 
=\left[\tilde{L_x}_1,\tilde{L_x}_2,\ldots,\tilde{L_x}_M\right]$, where $\tilde{\cdot}$ indicates a random draw from the Monte Carlo 
procedure for star $\nu$, we set to the data for which we want to compute the covariance matrix to be $X=\left[\textbf{L}_x^\nu,\textbf{L}_y^\nu,\textbf{L}_z^\nu\right]$. }. We then estimate the covariance matrix for each star, $\boldmatrix{L}$. To $\boldmatrix{L}$, we add the hyperparameters for each component, $\boldmatrix{H}^\kappa$, of which the $x$ and 
$y$ components are allowed to rotate via the additional hyperparameter $\gamma^\kappa$:
\begin{equation}
    \boldmatrix{H}^\kappa = 
    \begin{bmatrix}
    \sigma_{L_x^\kappa}\cos^2\gamma^\kappa + \sigma_{L_y^\kappa}\sin^2\gamma^\kappa                      & 
\left(\sigma_{L_x^\kappa}-\sigma_{L_y^\kappa}\right)\cos\gamma^\kappa\sin\gamma^\kappa & 0 \\
    \left(\sigma_{L_x^\kappa}-\sigma_{L_y^\kappa}\right)\cos\gamma^\kappa\sin\gamma^\kappa & \sigma_{L_x^\kappa}\sin^2\gamma^\kappa + 
\sigma_{L_y^\kappa}\cos^2\gamma^\kappa                       & 0\\
    0 & 0 & \sigma_{L_z^\kappa}                    
    \end{bmatrix}                  
\end{equation}
such that the total covariance matrix for each component is
\begin{equation}
    \boldmatrix{C}^\kappa = \boldmatrix{L} + \boldmatrix{H}^\kappa.
\end{equation}
Note that this is specific to the case where we are only allowing two dimensions to be covariant \cite[cf. eq. 3 of 
][]{Penarrubia2021}. This is a reasonable assumption owing to the geometry of the problem: the stars we select are confined to the stellar disc, and therefore the vertical behaviour is somewhat decoupled from the in-plane motions.
We proceed to compute the evidence of the model as in \citet{Penarrubia2021}, using \texttt{multinest} \citep{multinest}, a nested sampler which calculates the posterior distributions and evidence of the model. All parameters are taken from software examples\footnote{See  \url{https://github.com/rjw57/MultiNest/blob/master/example_gaussian/params.f90}} except for the number of model dimensions (in our case, 24 for the model parameters) and the maximum number of live points, which we set to be 400.
Given the posterior distributions, we can compute the probability of membership for a star in a given component by following equation~4 in \citet{Penarrubia2021}, generalised to compute membership in each of the $K$ components.
To estimate the median probability of membership as well as the uncertainty, we randomly select 20 samples from the posterior distribution and compute the median for each particle\footnote{In principle we could use the entire posterior distribution, however we compared the mean and scatter values with 20 random samples to the full distribution and found that the subset was a reasonable approximation of the full distribution.}. Throughout the work, we use samples from the posterior distributions directly to compute uncertainties, e.g. Figure~\ref{abun}. 

In Table~\ref{tab:fitparams}, we report the parameters of interest\footnote{While we fit in one set of parameters, it is more intuitive to report the Cartesian location of the centroid of each Gaussian $[L_x,L_y,L_z]=[\langle L\rangle \sin\theta\cos\phi,\langle L\rangle \sin\theta\sin\phi,\langle L\rangle \cos\theta]$, as well as the hyperparameters in units of degrees and $\kpc~\kms$. We also ensure that the fractions in each analysis are normalised properly.} for each of the components fit in the different analysed bins. We report the median of the posterior distribution as well as the 16$^{\rm th}$ and 84$^{\rm th}$ percentiles as uncertainties. Different runs are separated by horizontal lines. Bar, disc, and knot components are labeled. Unidentified components are reported, but are given no interpretation. In general, the components are centred close to $L_x=L_y=0$, with increasing $L_z$ magnitude as radius increases.

\section{Results}
\label{sec_results}
\subsection{Disentangling the inner Galaxy in chemistry-dynamics}

Figure~\ref{summary} displays the \textsl{\textsl{APOGEE}}-\textsl{Gaia} inner Galaxy sample in key kinematic diagrams. The top row shows the normalised angular momentum distribution of the parent sample weighted by the probability of stars being associated with the different components identified. The middle and bottom rows show these data grouped into the three components identified by our Gaussian mixture decomposition in angular momentum and position space, respectively. From the position diagrams (bottom row), the \textsl{\textsl{APOGEE}} pencil-beam selection function is clearly apparent. This complex footprint makes any characterisation of inner Galaxy populations based on position measurements alone very difficult, as one needs to account for the complex spatial selection function of the survey. This issue is largely overcome by analyzing the sample in angular momentum space, where these populations approximately follow Gaussian distributions (middle row). We run our method with two, three, and four Gaussian components, finding that three components are sufficient to describe the data. The three components identified are: bar (red), disc (green), and surprisingly, an additional spatially concentrated and non-rotating population, that we refer to in this work as ``knot'' (blue). All three components exist co-spatially at the centre of the Milky Way. The disc persists at all radii, emphasising that the inner Galaxy has a reservoir of stars that are not on bar-like orbits (see Table~\ref{tab:fitparams}).

\subsection{Inferring the origin and evolution of inner Galaxy populations using chemical abundances}
\label{sec_abundances}

\begin{figure*}
    \centering
    \includegraphics[width=1\textwidth]{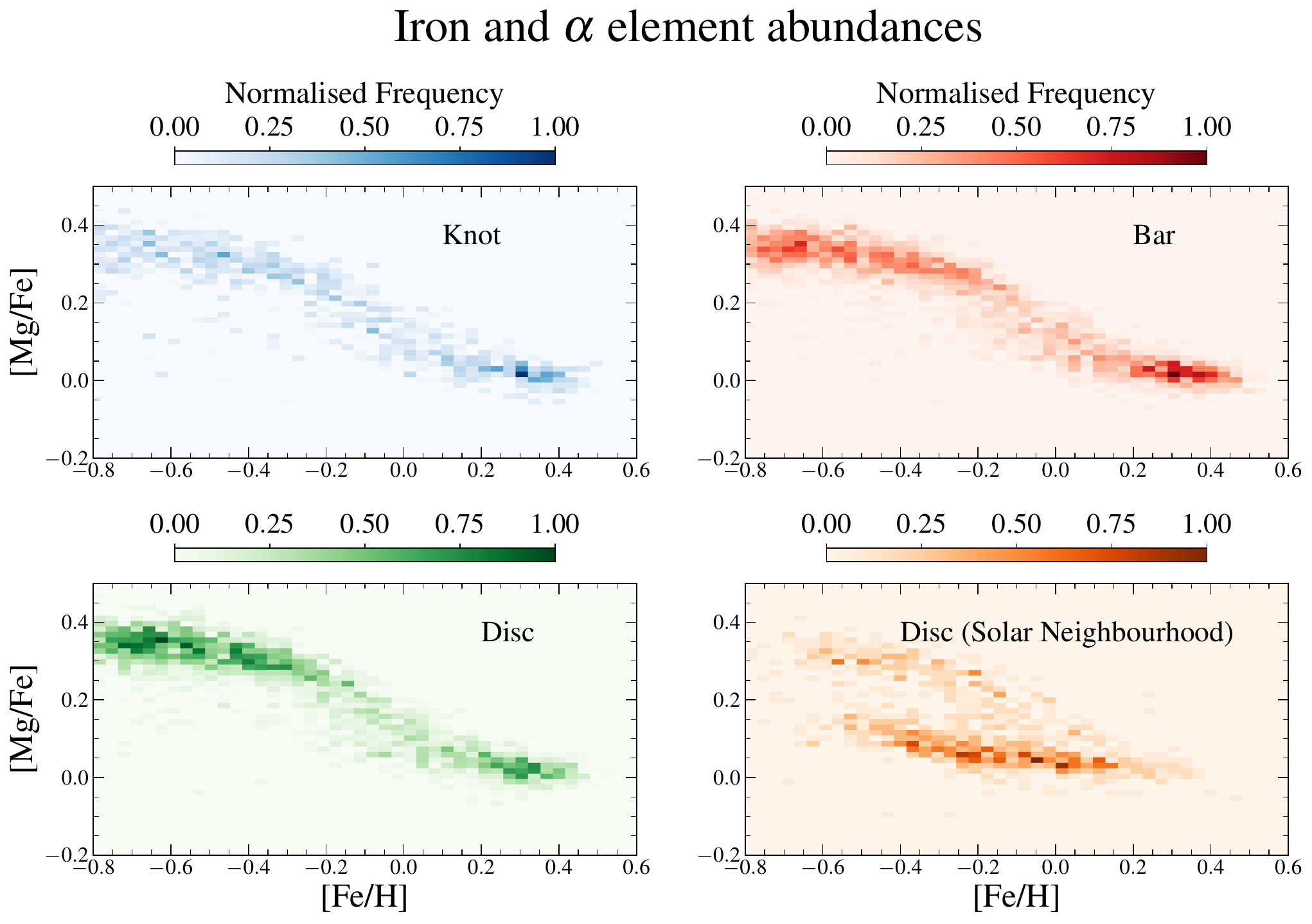}
    \caption{Median [Mg/Fe]-[Fe/H] density distributions for the knot (top left), bar (top right), and inner disc (bottom left) samples from 20 posterior chains of probability values computed using the angular momentum Gaussian mixture model decomposition procedure. Only stars within $R<3~\kpc$ and $1.5>\log~g>0.5$ are shown in the knot, bar, and inner disc panels, to account for systematic trends with surface gravity in our element abundance measurements. For reference, we also show disc populations in the solar neighbourhood (bottom right), defined the same way as the knot/bar/inner disc sample but for stars with heliocentric distances smaller than $d_{\odot}<2~\kpc$ (again, only showing stars with $1.5>\log~g>0.5$). The knot, bar, and inner disc samples follow a single track in the $\alpha$-Fe plane. All three populations show a trough in [Mg/Fe] at [Fe/H] $\sim$ solar, and are vastly different from disc populations in the solar neighbourhood; the solar neighbourhood sample also displays the known disc $\alpha$-Fe bimodality \citep{Hayden2015}, which is not seen for inner Galaxy populations. The inner disc sample shows a higher fraction of stars in the high-[Mg/Fe] sequence than the bar and knot (see also Fig~\ref{fig_hist}). The similar chemical abundances between the knot, bar, and inner disc suggests that these populations share a similar star formation history and likely formed over similar timescales. Thus, the chemical compositions of these three samples indicate that they are likely related. The different abundances between the inner and solar neighbourhood discs suggests different star formation histories. The higher [Fe/H] abundances of the inner disc suggests that it formed more rapidly and with a higher star formation rate/efficiency when compared to the disc in the solar neighbourhood, in contrast to recent studies \citep{Bovy2019}.}  
    \label{abun}
\end{figure*}

\begin{figure*}
    \centering
    \includegraphics[width=\textwidth]{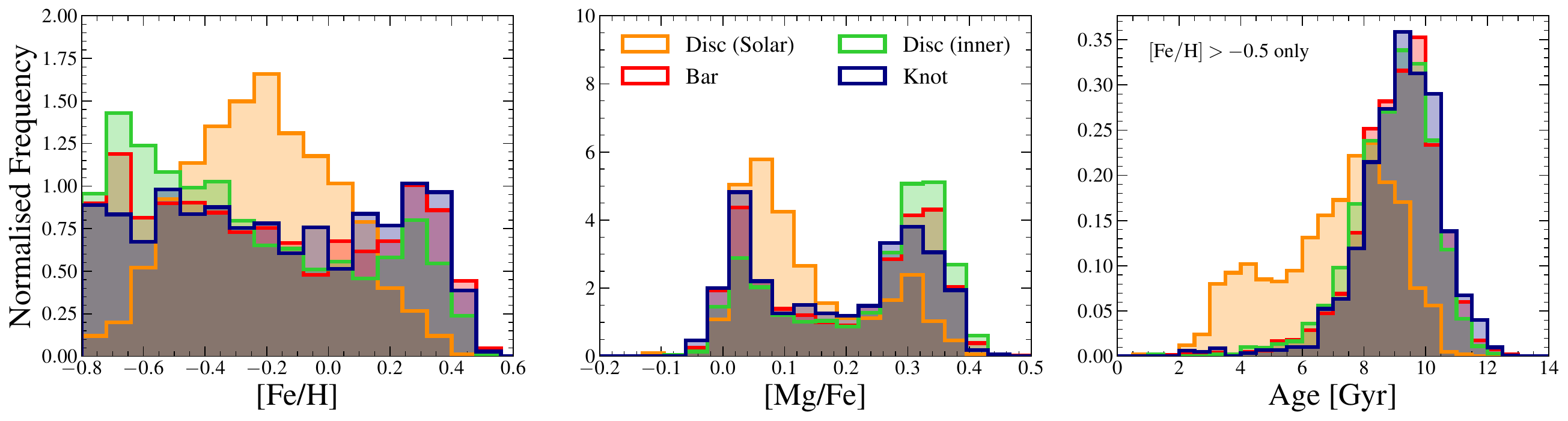}
    \caption{Normalised histogram distributions of [Fe/H] (\textit{left}), [Mg/Fe] (\textit{middle}), and age (\textit{right}) for stars in the knot, bar, inner disc, and solar neighbourhood disc. Only stars with $0.5<\log~g<1.5$ are used, to ensure no metallicity gradients and/or systematic effects in the abundances. Moreover, for the right panel, a cut of [Fe/H]$>-0.5$ is performed to ensure trustworthy ages, as below this metallicity there were no stars in the training sample used in \textsl{astroNN}. The inner disc sample is dominated by stars that are more [Fe/H]-poor and have higher [Mg/Fe] abundances when compared to the bar/knot. This is also the case when comparing the bar to the knot, although the differences are more subtle. In fact, there is a gradient in $\alpha$-Fe, whereby the distribution for the knot is weighted to lower [Mg/Fe], the bar sits in between the knot and inner disc, and the inner disc is weighted towards more enhanced [Mg/Fe] abundances. We find that the age distribution of the knot, bar, and inner disc are all qualitatively the same, and are comprised by old stars (Age $\gtrsim7$ Gyr). The abundances and ages of the solar neighbourhood disc are vastly different to the inner Galaxy populations, in contrast to recent findings using similar data \citep[][]{Bovy2019}. }  \label{fig_hist}
\end{figure*}

A useful way to infer the history of formation of Galactic populations is by examining the surface element abundance ratios of stars. This is because the element abundances largely remain invariant with time, allowing one to examine a fossilised record of the formation time and environment in which stars were formed \citep{Freeman2002}. Two of the most examined chemical abundances are $\alpha$ and Fe, as the combination of these trace the rate of core-collapse (SNII) and type-Ia supernovae (SNIa) production, two of the primary contributors to the chemical evolution of galaxies. 

Fig~\ref{abun} shows the [$\alpha$/Fe]-[Fe/H]\footnote{We use Mg as our $\alpha$ tracer element.} abundances of the knot, bar, and inner disc components, each weighted by the median of the probability value of every star being associated with a given component from 20 randomly-chosen posterior distribution samples of our fitting procedure. For reference, we also show the density distribution in this plane of the disc in the solar neighbourhood, defined in the same way as the parent sample (see Section~\ref{sec_data}), but for stars with heliocentric distances smaller than $d_{\odot}<2~\kpc$. For the knot, bar, and inner disc panels shown in Fig~\ref{abun}, we only use stars within $R<3~\kpc$ and that have a surface gravity range between $0.5<\log~g<1.5$, to minimise the effect from systematic uncertainties and abundance trends with stellar parameters \citep[e.g.,][]{Weinberg2021,Horta2023}. For the solar neighbourhood panel, we also limit the sample to stars with $0.5<\log~g<1.5$.

The knot, bar, and inner disc follow a single track in the $\mathrm{[\alpha/Fe]}-\mathrm{[Fe/H]}$ plane, indicating that they all share a common star formation history. All three samples appear in the high-$\alpha$ regime at sub-solar [Fe/H], and follow a trajectory that leads to a decrease in [$\alpha$/Fe] for super-solar metallicities (i.e., [Fe/H] $>0$), where the downward inflexion point (i.e., the $\alpha$-Fe ``knee'') occurs at $\mathrm{[Fe/H]}\sim-0.3$. However, the relative difference in the fraction of stars occupying the high-/low-$\alpha$ sequence in each sample is different, which is indicative that there could be a small age difference. Here, the inner disc contains more stars in the high-$\alpha$ sequence (i.e., has more stars that are less chemically evolved), than the bar. A similar conclusion can be reached from comparing the bar to the knot (i.e., the bar has more chemically unevolved stars than the knot). We attempt to quantify the relative differences in the number of stars in the high-/low-$\alpha$ sequences between the different samples in Fig~\ref{fig_hist}. Here, we also show the histogram of the age distribution, using the age measurements from the \textsl{astroNN} catalogue \citep{Leung2019}\footnote{We note that as the \textsl{astroNN} catalogue was trained on stars with asteroseismic information with [Fe/H] $>-0.5$. Thus, we only show those stars in our sample more metal-rich than this value. Moroever, we use these ages for examining qualitative differences only, and therefore should not be treated as accurate age distributions for these samples.}. As in Fig~\ref{abun}, we only use stars within $R<3~\kpc$ and $0.5<\log~g<1.5$ in Fig~\ref{fig_hist} for the knot, bar, and inner disc, and stars within $0.5<\log~g<1.5$ and $d_{\odot}<2~\kpc$ for the solar neighbourhood disc.

In summary, Fig~\ref{fig_hist} confirms our by-eye expectations from Fig~\ref{abun}, that there are more high-$\alpha$ and metal-poor ($-0.8<\mathrm{[Fe/H]}<0$) stars in the inner disc sample compared to the bar and knot, and more of these stars again in the bar when compared to the knot. However, the metallicity distribution function, and distribution in [Mg/Fe], is still quite similar for all samples, and spans the same range ($-0.8<\mathrm{[Fe/H]}<0.5$ and $-0.1<\mathrm{[Mg/Fe]}<0.4$, respectively). Interestingly, we find that when inspecting their age distributions, all three samples (knot, bar, and inner disc) present a similar age profile, which is old ($7<\mathrm{Age}<12$ Gyr). This result supports the hypothesis that all three samples share a similar star formation history and formed over a similar timescale. However, the different relative fraction of stars in the high-/low-$\alpha$ sequences between the three samples suggests some age difference. We argue that the reason we don't see any large age difference between the knot, bar, and inner disc samples is likely because any possible age difference is too subtle to be distinguished within the uncertainties on the reported age measurements. 

Moreover, when comparing the [$\alpha$/Fe], [Fe/H], and age distributions between inner Galaxy populations and the solar neighbourhood disc, we find that in all distributions these populations are vastly different. The solar neighbourhood disc is comprised by stars with more solar-like abundances, and presents the known $\alpha$-bimodality \citep{Hayden2015}\footnote{The $\alpha$-bimodality is a feature whereby at fixed [Fe/H], there are two clear separate sequences in [$\alpha$/Fe].}, that is not present in the inner Galaxy populations; the solar neighbourhood disc also reveals a much younger age distribution ($3\lesssim$ Age $\lesssim9$ Gyr) when compared to inner Galaxy populations. This result suggests that the inner Galaxy formed over a different timescale and with a different star formation history when compared to stellar populations in the solar neighbourhood.

Lastly, in addition to the $\alpha$ and Fe ratios, we examined the distribution of these populations in different element abundance planes provided by \textsl{APOGEE} (namely, C, N, Al, Si, Mn, Ni, not shown) probing different nucleosynthetic channels. We find that for all chemical abundance planes examined, the knot, bar, and inner disc populations always follow the same trajectory in chemical abundance space and occupy the same locii, that is different from the disc sample around the solar neighbourhood. Thus, our results from examining the $\alpha$-Fe plane remain consistent across all element abundances studied.

\subsection{The bar's pattern speed and corotation radius}
\label{subsec:patternspeed}

\begin{figure}
    \centering
    \includegraphics[width=1\columnwidth]{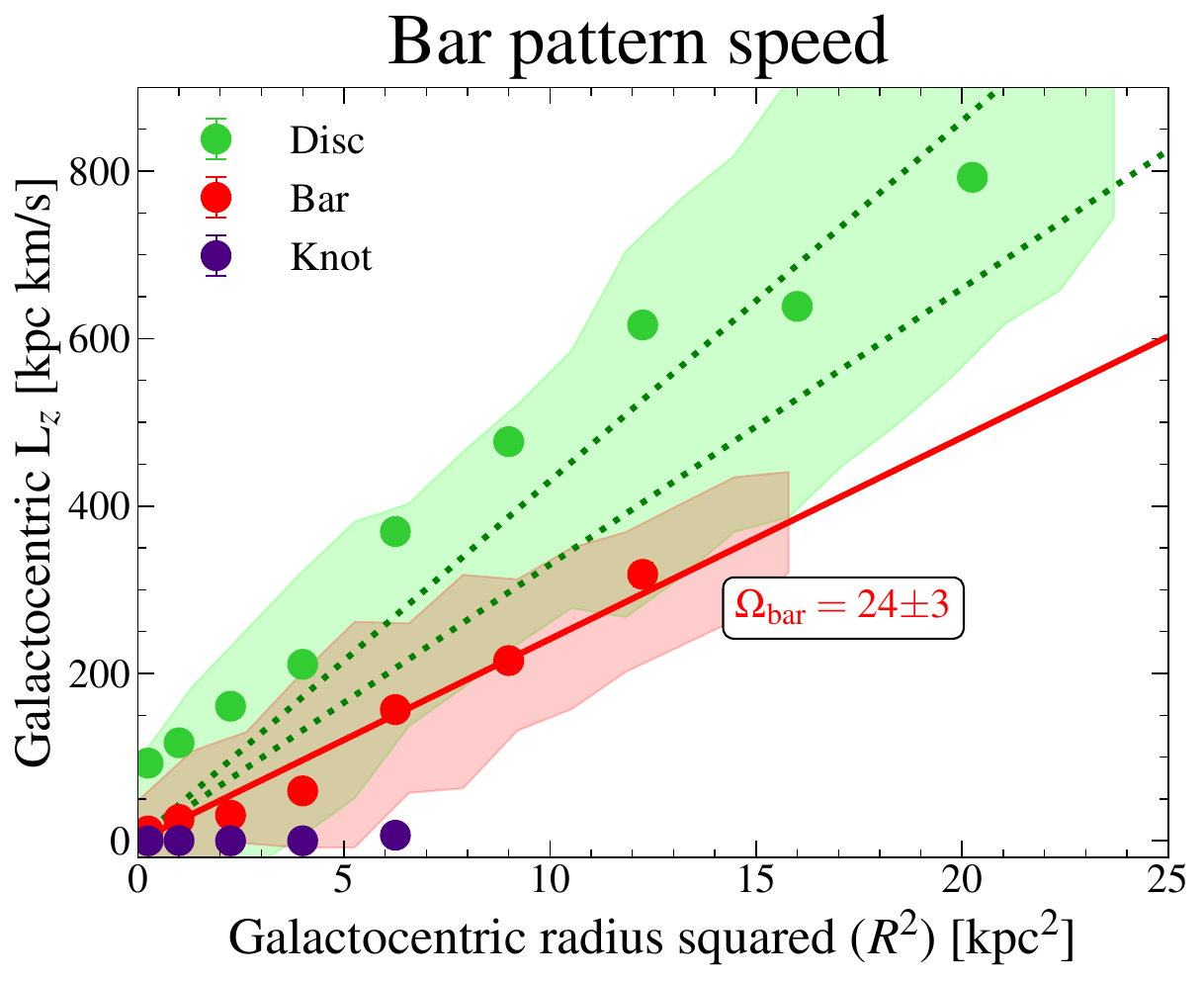}
    \caption{Azimuthal angular momentum as a function of Galactocentric radius squared. Markers show the values for the centroids of our model fits, whereas shaded regions show the 16$^{th}$ and 84$^{th}$ percentile range for the data with a probability of being associated to the bar (red) and inner disc (green) greater than $p > 0.8$. The red line is a linear regression fit to the fitted $L_{z}$ values for the bar component in independent radius bins, where the uncertainty is computed using the scatter in the corresponding $L_{z}$ values (i.e., $\sigma_{\mathrm{L}_{z}}$). We note that we also fitted the full data of bar member stars weighted by $p_{\mathrm{bar}}$ (i.e., a weighted fit to the bar sample), and found the same $\Omega_{\mathrm{bar}}$ value. The dashed green lines are linear models with slopes that match the high and low estimates of the pattern speed from previous works ($\Omega_{\mathrm{bar}}=[33,43]~\kms~\kpc^{-1}$). The fit to the bar sample yields a value for the pattern speed of $\Omega_{\mathrm{bar}} = 24\pm3~\kms~\kpc^{-1}$. This value is lower than recent estimates \citep{Portail2017,Bovy2019,Sanders2019}, which we find more closely agree with the inner disc data (green).}  \label{speed-angle}
\end{figure}

Figure~\ref{speed-angle} summarises the bar pattern speed measurement computed in this work (red line), determined by fitting a linear model to the bar component in the $L_{z}$-$R^{2}$ plane (accounting for the spread in $L_{z}$), where the slope is equal to $\Omega_{\mathrm{bar}}$. Our results yield a pattern speed for the Galactic bar of $\Omega_{\mathrm{bar}}=24\pm3~\kms~\kpc^{-1}$. This value is lower than the pattern speed found in previous works \citep[e.g.,][]{Portail2017,Bovy2019,Sanders2019}, which suggest a value between $\Omega_{\mathrm{bar}}\sim33-43 \mathrm{km~s^{-1}~kpc^{-1}}$. The difference is likely due to our ability to statistically distinguish between bar and disc components as a function of radius, rather than identifying the bar using $x-y$ map projections, which is bound to include contaminating untrapped stars from the disc, (in our analysis, stars with high $L_{z}$ are robustly assigned to the disc component).

\begin{figure*}
    \centering
    \includegraphics[width=\textwidth]{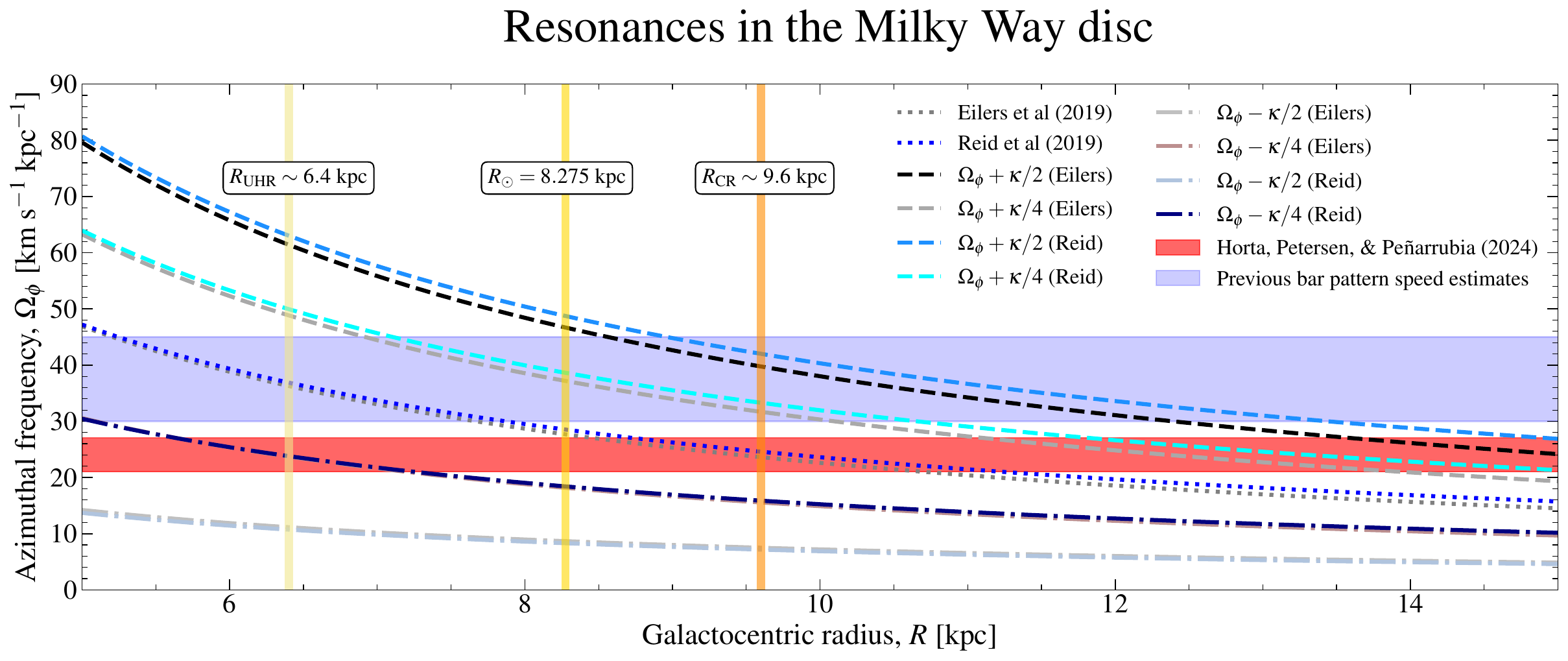}
    \caption{Azimuthal frequency as a function of (cylindrical) Galactocentric radius. Shown as dotted lines are the curves computed assuming the rotation curve of the Milky Way from \citet[][]{Eilers2019} (gray) and \citet[][]{Reid2019} (blue). As dashed (dashed-dotted) lines we also show the profiles of the positive (negative) resonance condition curves ($\Omega_{\mathrm{bar}} \pm \kappa/2$ or $\Omega_{\mathrm{bar}} \pm \kappa/4$). As a shaded red line we show the range of azimuthal frequencies we measure in this work for the Milky Way bar's pattern speed, $\Omega_{\mathrm{bar}}=24\pm3~\kms~\kpc^{-1}$. For comparison, as a blue shaded region we show the range in values for $\Omega_{\mathrm{bar}}$ reported in previous studies. The radius at which the dotted line crosses the value of the bar's pattern speed marks the corotation radius, which we estimate to be $R_{\mathrm{CR}}\sim9.4~\kpc$ using the \citet[][]{Eilers2019} rotation curve, and $R_{\mathrm{CR}}\sim9.8~\kpc$ using the \citet[][]{Reid2019} one; this leads to an average value of $R_{\mathrm{CR}}\sim9.6~\kpc$ (orange). This result implies that the solar neighbourhood is likely within the influence of the Galactic bar, and is in conflict with previous estimates, which suggest a corotation radius of $R_{\mathrm{CR}}\sim5-8~\kpc$. Moreover, the radius at which the $\Omega_{\mathrm{bar}} - \kappa/4$ profile crosses with the pattern speed demarks the ultra-harmonic resonance. We find this to be $R_{\mathrm{UHR}}\sim6.4$ for when assuming either of the two rotation curves used.}
    \label{fig_corotation}
\end{figure*}

Furthermore, pattern speeds in the range $\Omega_{\mathrm{bar}}=33-43~\kms~\kpc^{-1}$ (green-dotted lines in Fig~\ref{speed-angle}) fall between the pattern speed estimates of the disc and bar in our analysis (see the green dotted lines in Fig~\ref{speed-angle}), which may explain why previous analysis that do not dynamically distinguish between the bar and inner disc components find systematically higher values of $\Omega_{\rm bar}$. In fact, when determining the bar pattern speed in our mock sample by selecting bar stars in an $x-y$ map (i.e., a face-down projection of the Galactic disc), we recover a value of $\Omega_{\mathrm{bar}}$ that is biased high owing to inclusion of untrapped disc stars as well as sightline effects. 

Our estimate of $\Omega_{\mathrm{bar}}$ has large ramifications on our understanding of the dynamics of the Milky Way disc. For example, given the pattern speed of the bar we can compute the radial location of resonances, $\Omega_{\mathrm{bar}}=\Omega_{\phi}(R)-\kappa(R)/m$, where $\kappa$ is the radial frequency. The corotation radius, $R_{\mathrm{CR}}$, defined as the radius where the azimuthal frequency of the Milky Way disc is equal to the pattern speed of the bar ($\Omega_{\mathrm{bar}}=\Omega_{\phi}(R)$), is a key bar resonance. To measure this value, in Figure~\ref{fig_corotation} we show the value of constant azimuthal frequencies, $\Omega_{\phi}$, as a function of Galactocentric cylindrical radius, computed adopting the rotation curve from \cite{Eilers2019} and \citet{Reid2019}. This diagram is useful to determine the value of the corotation radius, as the radius in which the pattern speed of the bar crosses the profiles of azimuthal frequency correspond to the radius of corotation. Our value of the bar's pattern speed, $\Omega_{\mathrm{bar}}=24\pm3~\kms~\kpc^{-1}$, is shown in this diagram as a red shaded region. Using this value, we estimate a corotation radius of $R_{\mathrm{CR}}\sim9.4~\kpc$ using the \citet[][]{Eilers2019} rotation curve, and $R_{\mathrm{CR}}\sim9.8~\kpc$ using the \citet[][]{Reid2019} one; this leads to an average of $R_{\mathrm{CR}}\sim9.6~\kpc$, shown in orange. We also show in this diagram as a blue shaded region the range of previous pattern speed estimates, for comparison, which suggest corotation between $R_{\mathrm{CR}}\sim5-8~\kpc$. Our estimates of the pattern speed suggest that corotation (and by default the outer Lindblad resonances) are located at larger Galactocentric radii than previously thought, and that the solar neighbourhood is likely within the influence of the Galactic bar (the solar radius is shown as a vertical gold line).  

Along those lines, from Fig~\ref{fig_corotation} other important resonances caused by the Galactic bar can be measured. For example, the point in which the bar's pattern speed crosses with the profile of the negative resonance condition curve with $m=4$ (i.e., $\Omega_{\phi}(R)-\kappa(R)/4$) corresponds to the radius of the ultra-harmonic resonance. The ultra-harmonic resonance has been proposed to explain features in external galaxies such as rings \citep{Buta.1986} and `dark gaps' (\citealp{Krishnarao.2022}, though see also \citealp{Buta.2017} for a competing interpretation). Using our measurement of the Milky Way bar's pattern speed and assuming either of the two rotation curves used in this study, we estimate a radius of ultra-harmonic resonance of $R_{\mathrm{UHR}}\sim6.4~\kpc$ (see Table~\ref{tab:barparams}). 

Lastly, the ratio between the corotation radius and bar length, $\mathcal{R} = R_{\mathrm{CR}}/R_{\mathrm{bar}}$, is a measure of how much dynamical friction the halo exerts on the bar, and therefore the amount of dark matter in the inner Galaxy. Using the bar length measurement from \citet{Wegg2015} ($R_{\mathrm{bar}}=5.0\pm0.2~\kpc$), we estimate $\mathcal{R}=1.88^{+0.26}_{-0.2}$ and $\mathcal{R}=1.96^{+0.28}_{-0.22}$ using the rotation curve from \citet{Eilers2019} and \citet{Reid2019}, respectively. 

\begin{table}
\centering
  \begin{tabular}{ccc}
  \hline
  Parameter & Value \citep{Eilers2019} & Value \citep{Reid2019}\\
  \hline
  Pattern speed & \multicolumn{2}{c}{$24\pm3~\kms~\kpc^{-1}$}\\
  $R_{\rm UHR}$ & $6.4\pm0.8~\kpc$ & $6.4^{+1.0}_{0.7}~\kpc$\\
  $R_{\rm CR}$ & $9.4^{+1.3}_{-1.0}~\kpc$ & $9.8^{+1.4}_{-1.1}~\kpc$ \\
  $\mathcal{R}$$^{\rm a}$ & $1.88^{+0.26}_{-0.2}$ & $1.96^{+0.28}_{-0.22}$ \\
  \hline
  \multicolumn{3}{l}{$^{\rm a}$Assuming a bar length of $5.0\pm0.2$ \citep{Wegg2015}.}
  \end{tabular}
  \caption{Summary of bar parameters derived in this work. For completeness, we use rotation curve measurements from both \citet{Eilers2019} and \citet{Reid2019} to derive quantities.}
  \label{tab:barparams}
\end{table}

\section{Discussion}
\label{sec_discussion}

\subsection{The chemical compositions of inner Galaxy populations}
\label{sec_discussion_abun}

The similar trend seen in the high-/low-$\alpha$ sequences between the inner disc, bar, and knot (i.e., their track in chemical space, see Fig~\ref{abun}) indicates that these three populations share a common star formation history and that they likely formed over similar timescales; this result is confirmed when looking at their metallicity, [Mg/Fe], and age distributions (Fig~\ref{fig_hist}), as all three present a similar distribution and range. Thus, the chemical compositions of the knot, bar, and inner disc populations suggest that these are all likely related.  Conversely, stars in the disc near the solar neighbourhood present vastly different chemical abundances to inner Galaxy populations (see Section~\ref{sec_abundances} for details).

The chemical compositions of populations in the MW bulge region have been studied before using either large samples from stellar surveys or smaller samples with high-resolution spectroscopy \citep[e.g.,][]{Bensby2017,Barbuy2018,Zasowski2019,Hasselquist2020,Rojas2020,Queiroz2021,Recio2023,Guiglion2024}\footnote{Although we must also acknowledge the seminal work on chemical abundances in the inner Galaxy \citep[e.g.,][]{Whitford1986,Rich1988,McWilliam1994,Minniti1995}.}. These studies have shown that inner Galaxy populations present a wide metallicity range (spanning $-1.5\lesssim$[Fe/H]$\lesssim0.4$) and are typically old (Age $>8-10$ Gyr), although younger stars have also been found in this region  ($3\lesssim$ Age $\lesssim8$ Gyr: \citealp[e.g.,][]{Bensby2017}). These studies have also provided evidence for multiple episodes of significant star formation in this region, as well as the tentative presence of an $\alpha$-Fe ``knee'' at $\mathrm{[Fe/H]}\sim-0.3$ (for the high-$\alpha$ disc sequence). These observational results corroborate the findings presented in this work (Fig~\ref{abun} and Fig~\ref{fig_hist}). Moreover, these results have been backed by recent theoretical models, able to explain in large part the chemical abundance patterns and age profiles of inner Galaxy populations \citep[e.g.,][]{Debattista2023}.

One key feature we have found has been the trough in number density around $\mathrm{[Mg/Fe]}\sim0.1$ and $\mathrm{[Fe/H]}\sim0$ (i.e., solar) in the $\alpha$-Fe plane for the bar and inner disc samples \citep{Queiroz2021,Imig2023}, which is suggestive that there was a period in their evolution where star formation declined\footnote{The fact that these two samples show this similar feature is more evidence for these sharing a similar star formation history.}. \citet{Debattista2023} attributes this feature to a clumpy star formation history, and the fact that the low-[Fe/H] end of the high-$\alpha$ track is comprised by stars formed both \textit{in situ} and accreted via building blocks/accreted galaxies. This scenario falls in line with other observational findings \citep{Bensby2017}. However, these works also propose an age difference between stars on the low-$\alpha$ and high-$\alpha$ portions of the chemical track, which is not seen in our decomposition. This age difference is on the order of $\Delta\mathrm{Age}\sim7-10$ Gyrs (see for example Fig 20 of \citealt{Bensby2017}). We argue that these younger stars could be the result of contaminants from the Galaxy's disc, which as can be seen from the right panel of Fig~\ref{fig_hist} appear younger than inner Galaxy populations. While \citet{Bensby2017} provide good reasoning as to why the stars they observe are within the Galactic bulge, we postulate that it could still be possible that disc contaminants could be driving this age discrepancy, as other results using separate age estimates have shown inner Galaxy populations to be predominantly old \citep{Hasselquist2020}. While our chemical abundance results are agreement with previous work and suggest that the inner disc and bar had a period of lower star formation between the high-$\alpha$ and low-$\alpha$ sequences (i.e., the trough in Fig~\ref{abun}), we argue that the time delay between these two sequences cannot be greater than a few billion years, as the age distributions of the bar and inner disc samples are almost identical (see right panel of Fig~\ref{fig_hist}).

Furthermore, in this work we solely focused on looking at populations that belong to the disc and bulge ([Fe/H]$>-0.8$), in order to exclude metal-poor populations that likely belong to the innermost stellar halo. In doing so, we have not been able to identify a stellar halo component in the inner Galaxy, that would straddle the [Fe/H]-poor end of the metallicity distribution function. However, recent observational studies examining the chemical/dynamical properties of metal-poor stars in the inner Galaxy have provided evidence for the presence of such a population. For example, \citet{Horta2021} studied the chemical and dynamical properties of \textsl{APOGEE-Gaia} inner Galaxy stars and discovered what is likely to be the debris from a major building block (dubbed \textsl{Heracles}), for which counterpart globular clusters have been proposed \citep[e.g.,][]{Kruijssen2020,Forbes2020}. This population has been shown to be comprised by stars that are metal-poor ($-1.5\lesssim\mathrm{[Fe/H]}\lesssim-0.8$) and high-$\alpha$ ($\mathrm{[Mg/Fe]}\sim0.3$), and constitute some of the most chemically unevolved stellar populations in the Galaxy. This population is likely linked to and dates back to the proto-Milky Way (\citealp[][]{Belokurov2022,Rix2022,Horta2024_proto}). Moreover, in these inner regions is where you expect to find the largest contribution from evaporated/dissolved globular cluster stars (\citealp[][]{Schiavon2017_nrich,Trincado2019,Horta2021_nrich}), which would also contribute to the metal-poor tail of inner Galaxy populations. 

Previous works have suggested that the Galactic bar presents chemical compositions that resemble stars in the disc around the solar neighbourhood \citep[e.g.,][]{Bovy2019}. This result is surprising, as the Milky Way disc (and bar) is known to manifest different [$\alpha$/Fe]-[Fe/H] abundances at different Galactocentric radii \citep[e.g.,][]{Hayden2015,Imig2023}. Given these findings, it has been conjectured that the Galactic bar and solar neighbourhood disc formed following a similar star formation history, and over a similar timescale. However, our results from Fig~\ref{abun} and Fig~\ref{fig_hist} go against this hypothesis, as we find that disc populations around the Sun present vastly different chemical compositions when compared to inner Galaxy populations. We hypothesise that the reason for this discrepancy could be due to the different definitions used for a solar neighbourhood disc. For example, \citet{Bovy2019} defines the solar neighbourhood to be stars within $0.5<|z|<2~\kpc$ and $5<R<10~\kpc$, thus casting a wide net in radius across the Milky Way disc. This is a much more generous definition than the $d_{\odot}<2~\kpc$ one used in this study. While our results comparing inner Galaxy stars in and outside the bar corroborate the findings in previous studies, suggesting that these populations present the same chemical compositions, we argue that the solar neighbourhood disc and inner Galaxy do not present the same chemical abundance patterns. In turn, we conclude that the majority of stars in the solar neighbourhood disc did not form over a similar timescale and under a similar star formation history to the inner disc and bar.

\subsection{On the origin of the knot: a classical bulge component in the Milky Way's heart?}
\label{sec_discussion_knot}

Before speculating about a formation scenario for the newly identified knot population, we must begin by discussing why the knot does \textit{not} constitute existing components that are know to inhabit this region. Given the kinematic and dynamic properties of the knot, we argue that this population is unlikely to constitute a: 1) nuclear disc; 2) box/peanut/X-shaped bulge; 3) nuclear star cluster, for the following reasons: 1) the knot extends to $\sim1-1.5~\kpc$, whereas the nuclear disc extends to $\sim500~\pc$ \citep[e.g.,][]{Schultheis2021}. Moreover, the nuclear disc is assumed to be comprised by orbits with high in-plane (disc) rotation \citep[e.g.,][]{Sormani2022}, whereas the knot is made of stars on nearly-radial orbits with a spheroidal spatial distribution; 2) the box/peanut/X-shaped bulge is part of the bar, and thus shares the same angular momentum distribution as the bar \citep[e.g.,][]{Portail2015,Portail2015_banana,Debattista2017}, which is distinct from the knot (Fig~\ref{summary}); 3) the nuclear star cluster of the Milky Way is estimated to have a radius below $\lesssim0.5$ pc  \citep{Schodel2020}, whereas the knot extends to over $r\gtrsim1~\kpc$. 

Interestingly, the knot population appears to have the majority of the same characteristics a classical bulge would display \citep{Davies1983,Whitford1986,Wyse1997}. It is comprised by stars on radial orbits; it appears as a pressure supported spheroid in the central regions of the Galaxy; it is old; and it contains many stars with some of the highest (super-solar) [Fe/H] abundances in the Galaxy. However, as shown in Fig~\ref{fig_hist}, the metallicity distribution function of the knot spans metallicities down to [Fe/H]$\sim-0.8$, and is approximately flat across [Fe/H], which does not entirely fit the picture of a primarily super-solar abundance classical bulge \citep{Whitford1986, Rich1988}.

Classical bulges in $L^{\star}$ galaxies, as well as elliptical galaxies, are theorised to have formed primarily from major mergers \citep[e.g.,][]{Toomre1977,Barnes1992,Hopkins2010}. However, establishing if bulges in spiral galaxies form via the same framework as elliptical galaxies is still an open question \citep[e.g.,][]{Wyse1997}. If the classical bulge in the Milky Way formed via the theorised major merger channel, this would naturally imply that its constituent stars would be predominantly old and metal-poor (i.e., $\mathrm{[Fe/H]\lesssim-1}$), as we know that the Milky Way has had no major mergers since $z\gtrsim2$ \citep{Belokurov2018,Helmi2018,Haywood2018,Mackereth2019b,Myeong2019,Naidu2020,Horta2021}. This would contradict the chemical abundances found for the knot population (Fig~\ref{abun} and Fig~\ref{fig_hist}). In fact, the knot presents a metallicity distribution function that is approximately flat across $-0.8<\mathrm{[Fe/H]}<0.4$, and reaches some of the highest [Fe/H] values in the Galaxy. As discussed in Section~\ref{sec_discussion_abun}, there has been recent observational work that provides evidence for a metal-poor (and likely old) pressure supported system in the Galaxy's centre \citep{Horta2021,Rix2022}. This population would satisfy the classical bulge formation channel, and would also satisfy the majority of the conditions for a classical bulge (i.e., old, pressure-supported, spheroid, not forming stars), except for the metal-rich abundance condition. As noted in \citet{Wyse1997}, the condition of metal-rich abundances in a classical bulge is one that has been discussed extensively in the literature, and is still not fully established. Thus, 
we argue that the metal-poor spheroid found by recent observational work (i.e., \textsl{Heracles} \citealp[][]{Horta2021}, and/or the proto-Galaxy \citealp[][]{Belokurov2022,Rix2022}) is likely a more suitable candidate for this merger-formed type of classical bulge.

Another scenario for the formation of a (different type of) classical bulge in spiral galaxies is through secular scattering processes that heat stars from the disc/bar \citep{Spitzer.1951,Lacey.1984,Grand.2016,McCluskey.2024}. In this scenario, the resulting classical bulge formed by secular evolution would present very similar chemical abundances to the Galactic bar (the dominant population in the inner Milky Way), as found in this work. It would also appear as a pressure supported spheroid in the central regions of the Galaxy with stars on radial orbits, and if the inner disc/bar were formed by old stars, so would this type of classical bulge. However, a classical bulge formed via secular evolution would likely present a smaller velocity dispersion and would occupy a smaller volume than a merger-induced classical bulge.  These are exactly the characteristics found for the knot population. Moreover, depending on which stars in the bar get perturbed, and at what rate, it could also present a slight gradient in [Fe/H] and [$\alpha$/Fe] when compared to the bar (and inner disc), as found in this study (see Fig~\ref{fig_hist}).

We reason that the detection of the knot component in the central regions of the Galaxy provides evidence for a population that fits the mould of a classical bulge formed by secular evolution. This classical bulge is a pressure supported system (i.e., radial orbits; ${\bf L}\sim0~\kpc~\kms$), it is concentrated towards the innermost regions of the Galaxy ($r<1-1.5~\kpc$), it is old (Age $>8$ Gyr), and contains some of the most metal-rich stars in the Galaxy (up to $\mathrm{[Fe/H]}\sim0.4$). However, we note that this classical bulge definition is different to the one formed by major mergers \citep[e.g.,][]{Toomre1977,Barnes1992,Hopkins2010}, from which the elliptical galaxies are postulated to have formed. Thus, our results argue for an additional spheroidal component in the central regions of the Galaxy.

\subsection{Our measurement of Milky Way bar pattern speed in context}

Recent attempts to constrain the pattern speed of the MW bar have returned values in the range $\Omega_{\rm bar}\sim30-40  \mathrm{km~s^{-1}~kpc^{-1}}$ \citep{Portail2017,Sanders2019,Bovy2019,Binney2020}\footnote{This has typically been classified as the `slow' bar scenario in contrast to earlier works that found pattern speeds closer to $\Omega_{\rm bar}\sim50  \mathrm{km~s^{-1}~kpc^{-1}}$ \citep[e.g.][]{Dehnen2000}.}. The methods to estimate the bar pattern speed have split into two broad categories: (1) measuring the properties of bar stars themselves with an application of the continuity equation, and (2) measuring the effect of the bar at the solar neighbourhood via mapping of resonances. The former has the benefit of being a more direct tracer, while the latter technique benefits from a wealth of highly accurate data. Here, we have tried to chart a course that both uses excellent data, but in the region of the bar, coupled with a novel method. As our technique is more akin to the first method, we now explore the implications of our results for resonance-based measurements in the solar neighbourhood.

At present, it is clear that the solar neighbourhood is full of structure in the $v_\phi-R$ plane, such as the overdensities called `\textsl{Sirius}', `\textsl{Hercules}', `\textsl{Hyades}', and the `\textsl{Hat}'. What is not known is which bar resonances they may correspond to, if any\footnote{Other candidates, for example, would be resonances from spiral arms, or disequilibrium signatures \citep[e.g.][]{Hunt.2024}.}, though previous works have variously assigned features to bar resonances such as corotation, the outer Lindblad resonance, the 1:1 resonance, and the ultra-harmonic resonance(s) \citep[see, e.g.][]{Hunt.2019,Trick.2022,Wheeler.2022}. In fact, given a range of bar pattern speeds, \citet{Trick.2022} made predictions for resonant signatures, finding that ridges could be assigned to a range of resonances. In the scenario put forth by our pattern speed combined with rotation curve estimates, the `\textsl{Sirius}' feature in the solar neighbourhood would be possibly associated with the corotation resonance of the bar rather than a higher-order resonance \citep[cf. Figure 16 in][]{Gaia.anticentre.2021}.

Bars slow down over time owing to dynamical friction with the (dark) halo \citep{Tremaine.Weinberg.1984}, so a slow bar is not unexpected. Our pattern speed estimate suggests a slow bar, $\mathcal{R}\sim1.8-2$. Indeed, the bar in the model galaxy from \citet{Petersen.Weinberg.Katz.2021} analysed in Appendix~\ref{sec_mock} slows appreciably over the course of the simulation --- and is slowing at the time that the snapshot is taken. The slowing bar can have implications for where resonant features appear at a given time \citep{Chiba.2021a}. 

The implications of a bar with $\Omega_{\rm bar}=24 \mathrm{km~s^{-1}~kpc^{-1}}$ have not been explored in the literature. In particular, signatures in local structure both kinematic and chemical have not been studied for a solar neighborhood inside of the bar's corotation. Future interpretations of features in the solar neighborhood should allow for this possibility.

\subsection{Connection to external galaxies}

Approximately two thirds of local spiral galaxies host a bar in their central regions \citep{erwin2011}. Moreover, in the central regions of these extra-Galactic bars it is also typical to see additional spheroidal components \citep{Diaz2020,Fraser2020,Quejereta2021,Schinnerer2023}. Typically, these central components are hypothesised to be boxy/peanut/X-shaped bulges formed from the `buckling' of a bar, as may be true in the Milky Way \citep[e.g.,][]{Nataf2010,Wegg2013,Ness2016}. However, in this work we have provided evidence for an additional component (i.e., the knot), which given its chemical and dynamical properties, we suggest could comprise a classical bulge formed via secular evolution (see Section~\ref{sec_discussion_knot}). Recent theoretical work has suggested the formation of classical bulges via secular evolution and the destruction of short bars \citep{Guo2020}. We argue that studies of the shapes, kinematics, metallicities, and spatial extents of spheroidal components in the central regions of barred external galaxies could potentially help distinguish if these comprise a boxy/peanut/X-shaped bulge, a classical bulges formed via major mergers, or a classical bulges formed via secular evolution (like we propose for the knot in the MW).

\begin{figure}
    \centering
    \includegraphics[width=1\columnwidth]{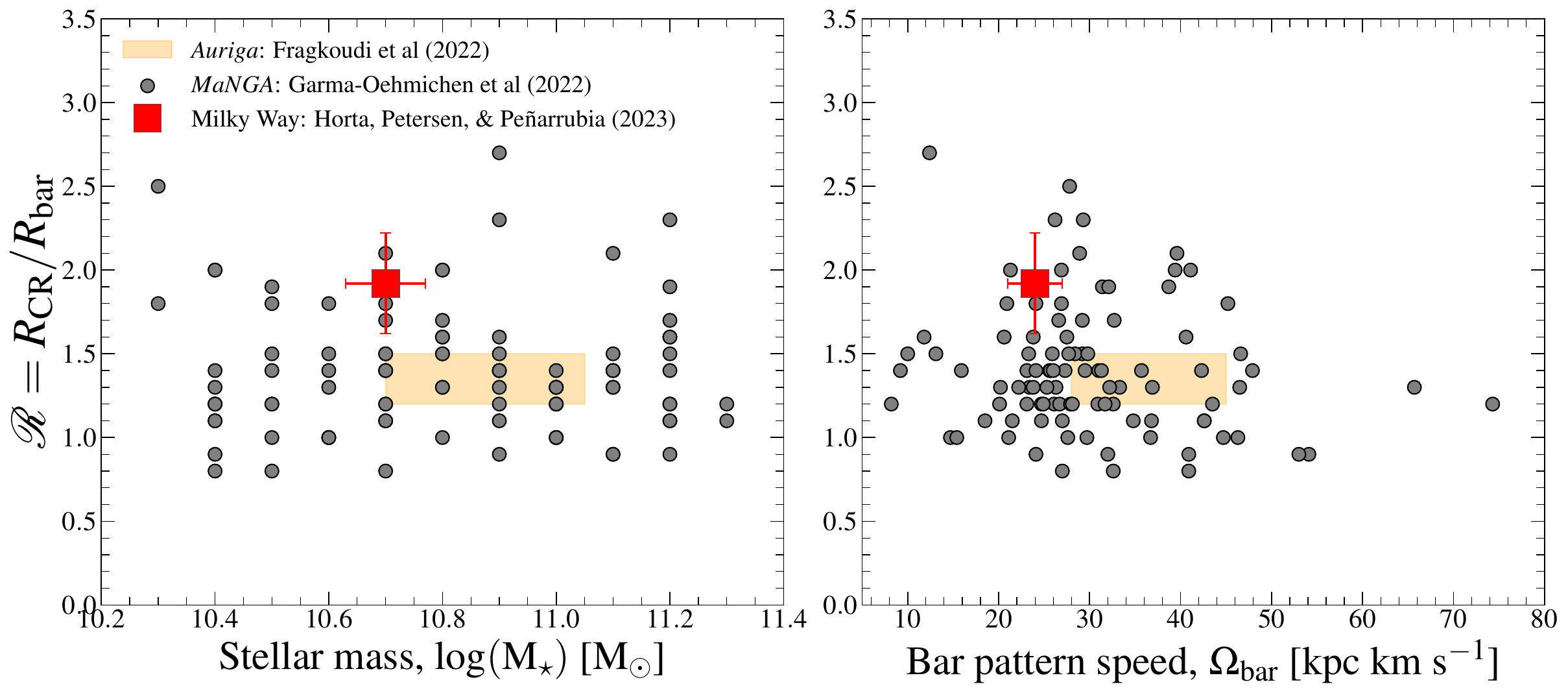}
    \caption{\textit{Left}: $\mathcal{R}$ (namely, $R_{\mathrm{CR}}/R_{\mathrm{R}}$) as a function of stellar mass for the Milky Way (red) derived in this work (see Table~\ref{tab:barparams}), assuming a stellar mass of M$_{\star}=5\pm1\times10^{10}$ [M$_{\odot}$] from \citet{Bland2016}, and Milky Way analogues in MaNGA \citep{GarmaOehmichen.2022}. For comparison we also show the average values found for $\mathcal{R}$ in the \textsl{Auriga} simulations \citep{Fragkoudi2021} in orange. The Milky Way sits slightly above the mean of the MaNGA galaxies and the average from the \textsl{Auriga} simulations. \textit{Right}: Same as \textit{left} but now for $\mathcal{R}$ as a function of bar pattern speed. Again, our estimates of the bar pattern speed and $\mathcal{R}$ place the Milky Way within the expected range of values for Milky Way-mass galaxies.}  \label{manga}
\end{figure}

Furthermore, in terms of the properties of bars, much work has been done trying to characterise the properties of bars in disc galaxies using integral-field unit (IFU) surveys and cosmological simulations.  \citet{GarmaOehmichen.2022} measured bar lengths and pattern speeds for a sample of 25 Milky Way analogues, finding a mean pattern speed of $\langle\Omega_{\rm bar}\rangle=30.48 ~\kms~\kpc^{-1}$, with mean corotation radius $\langle R_{\rm CR}\rangle =6.77 ~\kpc$. Taken with measurements of bar lengths, this corresponds to $\langle\mathcal{R}\rangle=1.45$. We show our value of $\mathcal{R}$ in comparison to these results in Fig~\ref{manga}. While the average value of $\mathcal{R}$ in the MaNGA galaxies \citet{GarmaOehmichen.2022} is smaller than our derived value (Table~\ref{tab:barparams}), it still sits within the expected range of a Milky Way-mass galaxy. Moreover, our higher estimate of $\mathcal{R}$ may be a result of the quiescent evolution of the Milky Way and significant bar slowdown over time. Moreover, the pattern speed estimate we obtain falls well within the average value of pattern speeds for Milky Way-analogues in MaNGA ($\langle\Omega_{\mathrm{bar, MaNGA}}\rangle=30\pm11~\kpc~\kms$).

Along the same lines, in Fig~\ref{manga} we also show the results from \citet{Fragkoudi2021}, who compared measurements of $\mathcal{R}$ between three cosmological simulations (namely, \textsl{Auriga, Illustris, and EAGLE}), and found a wide range of mean values for $\mathcal{R}$ in barred galaxies. The values found ranged from $\mathcal{R}\approx1.2-1.5$ in \textsl{Auriga} to $\mathcal{R}\approx2.5-4$ in \textsl{EAGLE} and \textsl{Illustris}. Our pattern speed measurement leads to an inferred value of $\mathcal{R}\sim2$ (assuming a bar length of $5\pm0.2~\kpc$ \citealp[][]{Wegg2015}), which falls between the two discrepant results.

\subsection{Limitations in our method}

We apply an initial metallicity cut to the \textsl{APOGEE} sample ($\mathrm{[Fe/H]}>-0.8$) which removes the metal-poor stellar halo component. We find the results are not sensitive to small changes in the location of the cut, but when including the entire metallicity range of the \textsl{APOGEE} sample, our fits would require additional component(s). As we are not presently studying the stellar halo in the inner Galaxy, we find this metallicity cut to be an efficient selection to primarily select stars that form part of the dominant populations in this region, the inner disc and bar. 

Furthermore, our method approximates the angular momentum distribution of each component in a 1 kpc wide bin as a multivariate Gaussian. If the distribution departs significantly from Gaussianity, our method will incur biases in the characterisation of the different components. In tests with the mock galaxy sample, we find that locally, i.e. within 1 kpc wide bins, a multivariate Gaussian is generally a appropriate approximation. Of particular relevance to our study, we do not find any evidence that non-Gaussianity can create the `knot' component: in tests with the mock galaxy sample, when a component becomes non-Gaussian, the GMM decomposition will break up the component, but neither is centred at ${\bf L}=(0,0,0)~\kpc~\kms$. Moreover, using our mock galaxy we have tested the detection of the knot component by artificially injecting a knot and removing it, to see if our method can robustly identify an additional component when present. We find that we cleanly recover the artificial knot when injected. Thus, given these tests, we argue that any detection of a separate component using our angular momentum decomposition method should in fact reflect a true additional component.

\section{Conclusions}
\label{sec_conclusions}

The results presented in this paper can be framed into a bigger picture of galaxy formation and evolution of the Milky Way via the following summary:
\begin{itemize}
    \item The inner $\sim$5 kpc region of the Galaxy is complex. Using the photo-astro-spectroscopic data supplied by the \textsl{APOGEE-Gaia} surveys, we have applied a Gaussian Mixture Model to inner Galaxy stars in angular momentum space. Using our statistical decomposition, we have found that the inner Galaxy (above $\mathrm{[Fe/H]}>-0.8$) can be described by three components: an inner disc, a bar, and a previously unknown compact, non-rotating ($\langle \mathrm{L}\rangle\sim 0~\kpc~\kms$) component in the heart of the Galaxy with a spheroidal shape that we refer to as the ``knot'' (Fig~\ref{summary}).
    \item  The newly identified knot component appears to be different to a nuclear disc, a nuclear star cluster, and box/peanut/X-shaped bulge on account of its chemical-dynamical properties. However, its chemical-dynamic properties align with the definition of a classical bulge formed via secular evolution. Thus, our finding of the knot population provides potential evidence for a classical bulge in the Milky Way that is not formed by the theorised major merger channel, like that postulated for elliptical galaxies. The detection of the knot therefore adds an additional population to the list of components inhabiting the inner Galactic regions. It also adds an extra layer of complexity to the already puzzling inner Galaxy.
    \item The chemical compositions (especially, the [Mg/Fe]-[Fe/H] ratios) of this knot population are qualitatively the same as the bar and inner disc components, which suggests it likely formed under a similar star formation history and over a similar timescale (see Fig~\ref{abun} and Fig~\ref{fig_hist}). All three samples appear in the high-$\alpha$ regime at sub-solar [Fe/H], and follow a trajectory that leads to a decrease in [Mg/Fe] for super-solar metallicities (i.e., [Fe/H] $>0$), with a downward inflexion point (i.e., the $\alpha$-Fe ``knee'') at $\mathrm{[Fe/H]}\sim-0.3$. Moreover, all inner Galaxy populations display vastly different chemical abundance and age distributions when compared to disc populations around the solar neighbourhood. 
    \item Using probabilistic estimates for Galactic bar membership from our angular momentum decomposition, we have estimated a fresh measurement of the bar pattern speed. We find that the bar is more slowly rotating than previously reported ($\Omega_{\mathrm{bar}} = 24\pm3~\kms~\kpc^{-1}$, see Fig~\ref{speed-angle}; and $\mathcal{R}\sim1.9$ see Fig~\ref{manga}). With this new pattern speed measurement, we estimate that the influence of the bar extends outside the solar radius, leading to a corotation radius beyond the Sun, at $R_{\mathrm{CR}}\sim9.6~\kpc$ (Fig~\ref{fig_corotation}).
\end{itemize}

The future of studies aiming to examine the intricate properties of stellar populations in the Milky Way's centre looks promising. Large stellar surveys are currently delivering exquisit astrometric, photometric, and spectroscopic information for large samples of stars in this region. This data will only grow exponentially with the upcoming SDSS-V \citep{Kollmeier2017}, \textsl{MOONS} \citep{moons}, and \textsl{4-MOST} \citep{DeJong2019} surveys. These data, in combination with detailed extra-Galactic observations of barred disc galaxies (e.g., \textsl{MANGA}: \citealp{Bundy2015}; \textsl{PHANGS-MUSE}: \citealp{Emsellem2022}; \textsl{SAMI}: \citealp{Bryant2015}; \textsl{JWST}: \citealp{jwst2006}) and theoretical predictions from cosmological simulations of Milky Way-mass galaxies (e.g., \textsl{Auriga}: \citealp{Grand2017}; \textsl{FIRE} \citealp{Hopkins2018,Hopkins2023}; \textsl{TNG50}: \citealp{Pillepich2023}) will help place constraints on the formation, evolution, and properties of the Galactic bar, and the formation of bulges in the central regions of Milky Way-mass galaxies.

\section*{Acknowledgements}
DH thanks Melissa Ness, Francesca Fragkoudi, Kathryn Johnston, and Martin Weinberg for helpful discussions, and Sergey Koposov for kindly hosting him at the Institute for Astronomy. He also thanks Sue, Alex, and Debra for their constant support. MSP acknowledges a UKRI Stephen Hawking Fellowship. The Flatiron Institute is funded by the Simons Foundation. This work has made use of data from the European Space Agency (ESA) mission
{\it Gaia} (\url{https://www.cosmos.esa.int/gaia}), processed by the {\it Gaia}
Data Processing and Analysis Consortium (DPAC,
\url{https://www.cosmos.esa.int/web/gaia/dpac/consortium}). Funding for the DPAC
has been provided by national institutions, in particular the institutions
participating in the {\it Gaia} Multilateral Agreement.

\section*{Data Availability}
All \textsl{APOGEE} and \textsl{Gaia} data used in this study are publicly available and can be downloaded directly from \url{https://www.sdss4.org/dr17/} and \url{https://gea.esac.esa.int/archive/}, respectively. Classifications for all stars as well as mock galaxy tests are available at \url{https://github.com/michael-petersen/BarGMM/tree/main}.

\section*{Software}
    Matplotlib \citep{Hunter:2007},
    numpy \citep{numpy},
    \texttt{multinest} \citep{multinest}.
    \texttt{EXP} \citep{Petersen.Weinberg.Katz.2022}.


\bibliographystyle{mnras}
\bibliography{refs} 



\appendix

\section{Tests on mock catalogues of barred galaxies}
\label{sec_mock}

To validate our statistical approach and solidify our dynamical understanding of the observations, we construct a mock \textsl{\textsl{APOGEE}} catalogue for which we know the input model parameters. We use a self-consistent \textit{N}-body simulation of a realistic Milky Way-mass galaxy \citep{Petersen.Weinberg.Katz.2021}. We emphasise, however, that the simulation is not a model of the Milky Way, but rather a generic disc galaxy with a bar and a similar stellar-to-halo mass ratio as the Milky Way.

To construct the mock, we draw from the relevant features of \textsl{\textsl{APOGEE}}. \textsl{\textsl{APOGEE}} is a pencil-beam survey. The coverage on the sky is therefore non-continuous. Coupled with a (roughly) uniform depth per sightline, this produces a unique pattern in 3D. Our goal in constructing mocks is to model the relevant features of the data distribution to understand any possible biases in our analysis due to the spatial selection function of the data. 

\begin{figure*}
    \centering
    \includegraphics[width=0.9\textwidth]{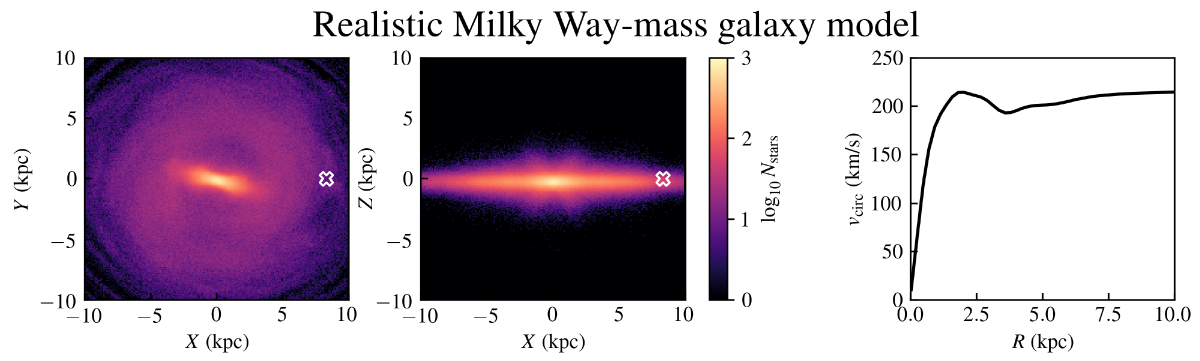}
    \caption{The model galaxy used to build the \textsl{\textsl{APOGEE}} mock. For the model galaxy, we show the face-on view (left panel), the edge-on view (middle panel), and the circular velocity curve (right panel). In the two left panels, we mark the location of the sun with a white `X'. The face-on panel shows the tested rotation angle, $12^\circ$ relative to the solar location. The edge-on panel shows that the model bar has formed an elevated population as well as some additional central concentration. The circular velocity is computed along the bar major axis.} 
    \label{fig:mock-bar-rotation}
\end{figure*}

We start with a model stellar disc that has self-consistently formed a bar. The simulated model starts as an initially bar-less disc with a similar stellar-to-halo mass ratio, as well as a rotation curve approximating that of the Milky Way. The simulation, evolved with {\sc exp} \citep{Petersen.Weinberg.Katz.2022}, forms a bar comprised of stars {\it trapped} into a self-reinforcing potential. The bar stars are drawn from the disc --- a disc component remains persistent throughout the evolution at all radii --- and the bar slows over time through resonant coupling with the halo. It is from this model that we sample particles to create a mock data set. As the model is originally in virial units, we scale the units to be in lengths of $\kpc$ and velocities of $\kms$, such that the scale length of the disc is $3~\kpc$, and the circular velocity at the Sun's position $(8.3,0,0)~\kpc$ is $V_\odot=232~\kms$. For the mock, we assume a simplified peculiar motion at the Sun's location of $\vec{v}_\odot=(9.,12.,9.)~\kms$, approximately equal to the observed estimates \citep{Schonrich2010}. Figure~\ref{fig:mock-bar-rotation} shows basic views of the model galaxy, including face-on and edge-on views, as well as the rotation curve. Broadly speaking, the model is a good match to the reported Milky Way parameters, though we stress it is not designed to be a Milky Way model.

The model has been extensively analysed \citep{Petersen.Weinberg.Katz.2021}, which provides significant benefit in the form of known classification of bar and disc stars, as well as global bar parameters. We therefore select a snapshot from the model that resembles our anticipated results for the Milky Way data, with $\Omega_{\rm bar}=26~\kms~\kpc^{-1}$. We set the angle of the end of the bar (as defined by the members of the bar-parenting family of orbits) relative to an observer at the solar location to $\theta_{\rm bar}=12^\circ$.

To select particles for inclusion in the mock, we first create a grid of sightlines with $\ell\in[-35.5^\circ,35.5^\circ]$, $d\ell=5.5^\circ$ and $b\in[-16^\circ,16^\circ]$, $db=5.5^\circ$, starting from $(\ell,b)=(0,0)$ to mimic the distribution seen in \textsl{\textsl{APOGEE}}. To avoid strongly sampling exactly in the plane, we offset the $b$ coordinates of the sightlines by $+0.5^\circ$. We define a circle around each point with radius $R=1.4^\circ$, calling each one of these a pencil beam. We avoid the centre of the galaxy by removing any sightlines that fall within a $R=6^\circ$ circle of $(\ell,b)=(0^\circ,0^\circ)$. The resulting on-sky distribution resembles the \textsl{\textsl{APOGEE}} sky coverage reasonably well, and although it is not an exact replica, it is sufficient for our purposes of investigating whether non-continuous coverage affects the results. We filter the model by retaining only the stars that appear within the mock sightlines observed from the Sun, at $(8.3,0,0)$ kpc.

The relative number of stars in each annulus radiating from the Galactic centre is a critical number to match between the data and the mocks. We achieve this approximate match by precisely matching the distance distribution of the input data set (from the Sun's modeled location). To match the distance distribution of an input data set, we first create a cumulative distribution function (CDF) of the distances, $n(r)$, normalised to 1. Inverting this curve, when then draw random variates between 0 and 1, which are mapped to the corresponding source distance. We then search the model for a particle within some tolerance of the desired distance, in this case, $3\%$. In practice, we find that the resulting cumulative distribution of distances in the mock data set matches the cumulative distribution of the \textsl{\textsl{APOGEE}} distances at the $1\%$ level. Our mocks are half the size of the \textsl{\textsl{APOGEE}} data, which makes the mocks a more stringent test of the method, as tests with larger samples demonstrate that parameter uncertainties decrease, as expected.

To ``observe'' the model, we create uncertainty functions by examining the uncertainty distributions in the \textsl{\textsl{APOGEE}} selection. We assume no uncertainties on the sky position of sources. We find that distance uncertainties show a strong linear dependence on distance, which we approximate as $\varepsilon_d = \mathcal{N}\left(0.17d-0.2,0.15\right)~\kpc$, while
proper motion uncertainties do not show a dependence on distance: $\varepsilon_{\mu_\ell,\mu_b} = \mathcal{N}\left(0.05,0.012\right)~\mathrm{mas~yr^{-1}}$. We model the correlation coefficient as a normal distribution centred on zero and a width approximately similar to the \textsl{Gaia} measurements:
$\rho_{\mu_\ell,\mu_b} = \mathcal{N}\left(0.,0.1\right)$.
Line-of-sight velocities do not show a strong dependence on distance, and we model them as $\varepsilon_{v_{\rm los}} = \mathcal{N}\left(0.07,0.02\right)~\kms$. When combined, these uncertainty models produce a qualitative match to the distributions in \textsl{\textsl{APOGEE}}. Additionally, we find that the mock data set has quantitatively similar angular momentum uncertainties (which are the important metric to match in our analysis), of order $\epsilon_{L_{\rm x,y,z}}\simeq100~\kpc~\kms$ per star.

We run the same fitting procedure on the mock data set, partitioning the data into eight bins of 1 kpc width, starting from $R=0$ kpc and increasing by 0.5 kpc (that is, the bins overlap such that we have four independent bins). In the discussion below, we denote the bins as $[r_{\mathrm{min}},r_{\mathrm{max}}]$. In each bin, we decompose the structure into three Gaussians in angular momentum distributions. 
We expect three kinematic components in the fits to the mock: the underlying disc (which is co-spatial with the bar), the bar, and a concentrated component (i.e., a knot) that arises from stars who have been robbed of their angular momentum by the bar through fluctuations in the potential. However, the components will not be present in all bins. As we always fit three components, this means that not all components may be significant, and may have associated low membership fractions $f$ (i.e. 1\% or less). Another possible analysis feature we identified through the mocks is the artificial splitting of one true component into two Gaussians, owing to non-Gaussianity in the angular momentum distributions resulting from the footprint. When this occurs, we combine the fitted Gaussians into a single component, for which the mean and dispersions may be analytically calculated (see for example bin $2-3~\kpc$ in Fig.~\ref{fig:mock-bar-pattern-speed}, which is the combination of two components). We find that the disc component is split in two of the eight bins of the mock sample, but do not find any evidence for `splitting' in the data fits. 

We identify three distinct components in the mock in the [0,1], [0.5,1.5], [1,2], and [1.5,2.5] kpc bins: knot, disc, and bar. The presence of a knot-like component is consistent with the evolution of simulated bars \citep[i.e.,][]{Debattista2017}.
As expected, we identify only two distinct components in the [2,3], [2.5,2.5], and [3,4] kpc bins: disc and bar. We identify only a disc component in the [3.5,4.5] kpc bin. The bar extends to $<3.5$ kpc in the model \citep{Petersen.Weinberg.Katz.2021}, which we recover in our Gaussian Mixture classifications.

Comparing with the classifications of bar stars in \cite{Petersen.Weinberg.Katz.2021}, we find that the GMM decomposition predominantly classifies high-probability member particles correctly. The primary uncertainty
modes are stars that lie near the overlap region of two Gaussians, e.g. disc stars with low instantaneous $L_z$ being confused for bar stars, or bar stars with low instantaneous $L_z$ being confused for knot stars. These sources of uncertainty are expected, and the classifications are still statistically robust, even if some individual particle classifications are uncertain in overlapping regions of angular momentum. Comparison of the average population properties suggests that the primary uncertain mode is true bar stars being classified as disc stars at small radii. This is unavoidable given (1) current angular momentum uncertainties, which are comparable to the angular momentum shift between these components and (2) the inherent overlap in angular momentum distributions at small radii. Future data sets with smaller uncertainties will result in more robust classifications of the different populations. However, we stress that the classification ambiguity between the bar and disc stars at small radii does not affect our inferences of global properties of the components: e.g. the pattern speed of the bar and the angle of the end of the bar are well-recovered (see next paragraph). However, the inner bins, at $R<2$ kpc, show some biases in $\gamma_{\mathrm{bar}}$ as a result of the aforementioned angular momentum overlap, exacerbated in some cases by the footprint coverage of \textsl{\textsl{APOGEE}} (see below). 

\begin{figure*}
    \centering
    \includegraphics[width=0.9\textwidth]{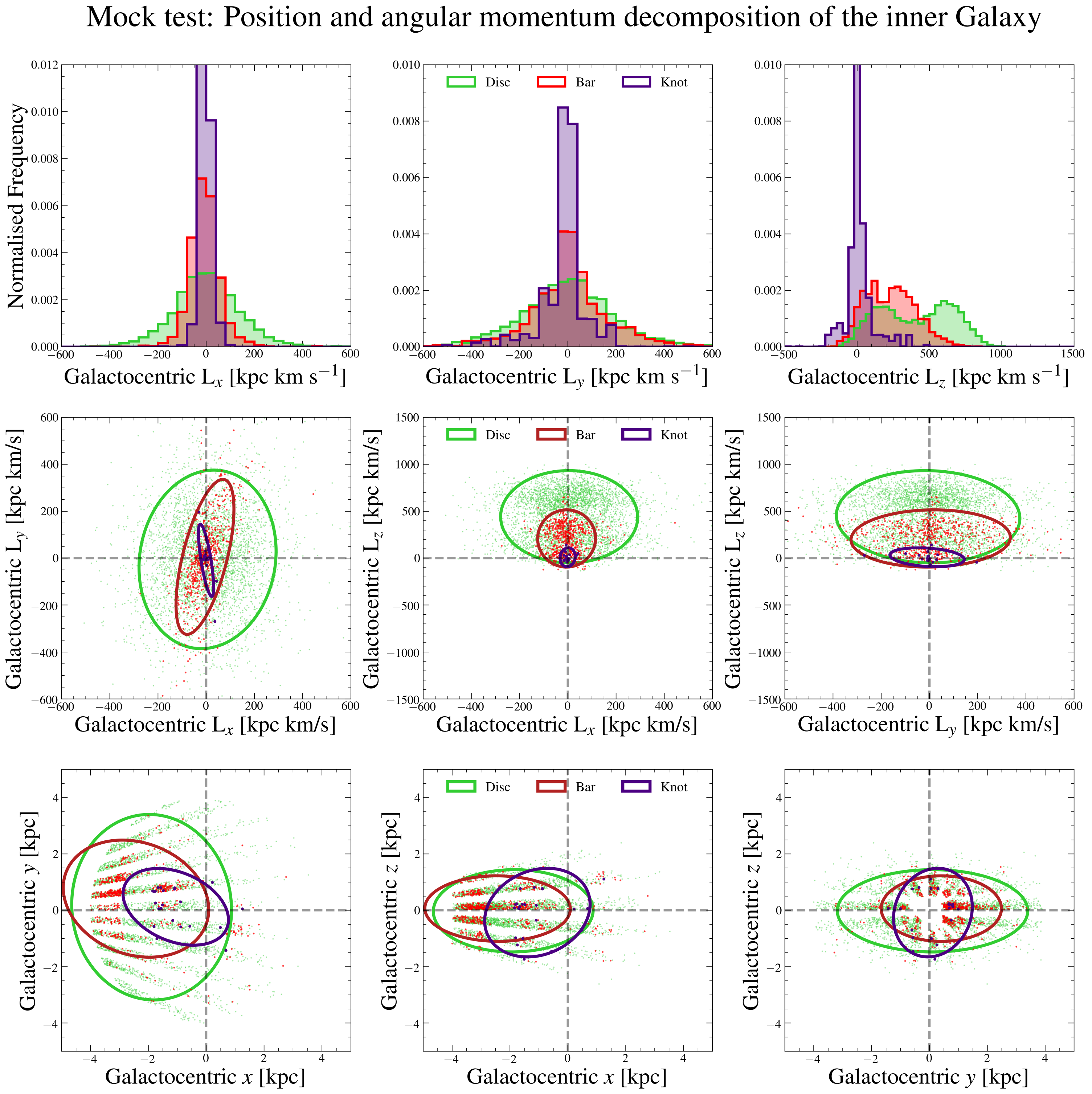}
    \caption{Angular momentum and spatial distribution summary for the mock {\sl \textsl{APOGEE}} data set. The data set is described in the Section~\ref{sec_method}. The layout of the panels is the same as Figure~\ref{summary}. In general, the mock data set shows the same features as the real data, though the knot component is a smaller fraction.} 
    \label{fig:mock-bar-summary}
\end{figure*}

\begin{figure}
    \centering
    \includegraphics[width=1\columnwidth]{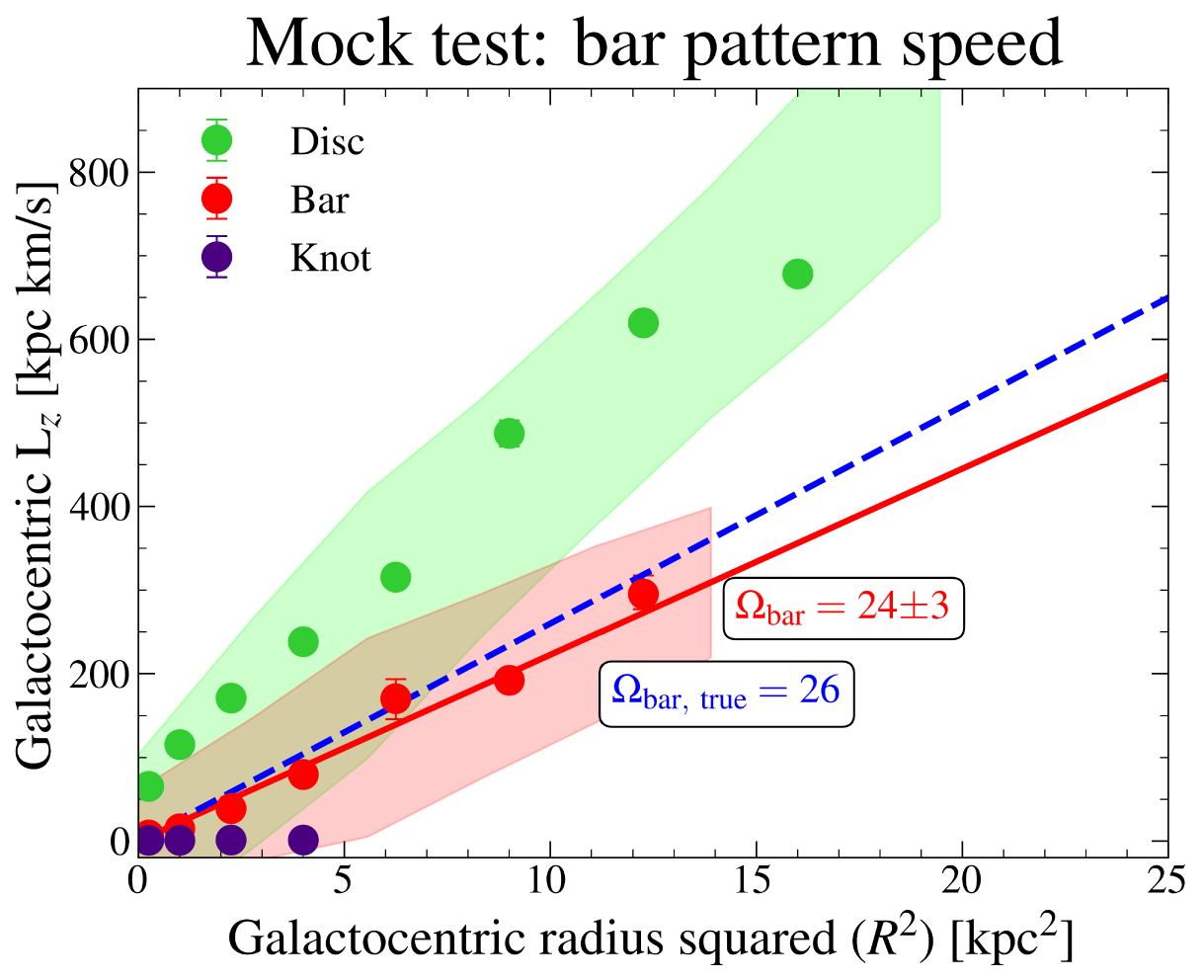}
    \caption{Measurements of the pattern speed of the bar in the mock {\sl \textsl{APOGEE}} data set. The layout is the same as Figure~\ref{speed-angle}. Given that we know the true values for the mock data set, we are able to compare the fit parameters with the true values.}
    \label{fig:mock-bar-pattern-speed}
\end{figure}

We stress that owing to the limited footprint of the observations we cannot measure the overall fraction of stars in different components, but the fraction of that component {\it in the footprint}.
In Figure~\ref{fig:mock-bar-summary}, we make the analogous plot to Figure~\ref{summary} for the mock data set. We show the angular momentum summary for the three components identified by the GMM, which follow very similar distributions to the real data. The primary differences owe to the small number of stars in the [0,1] kpc bin in the mock ($N=381$) versus the data ($N=1429$), which is a result of the footprint mismatch with the {\sl \textsl{APOGEE}} data. However, this difference does not appear to affect the classifier, which still robustly identifies three distinct components.
In Figure~\ref{fig:mock-bar-pattern-speed}, we make the analogous plot to Figure~\ref{speed-angle} for the mock data set. The correspondence between the model and mock is robust, and demonstrates that the mock is capturing much of the salient dynamics in the real Milky Way.
Using the same methodology as in the main text, we measure $\Omega_{\rm bar}=24\pm3~\kms~\kpc^{-1}$. When compared to the true value ($\Omega_{\rm bar}=26~\kms~\kpc^{-1}$), we find that we recover the pattern speed of the bar within measurement uncertainties.

We also construct a second mock using the same procedure as above, {\it except} we do not apply any pencil-beam footprint constraints, instead only confining the selection to particles with $\ell\in[-35.5^\circ,35.5^\circ]$. That is, we begin with the model galaxy and impose only the cumulative distance distribution of our {\sl \textsl{APOGEE}} data set to select particles for the footprint-free mock. This mock is designed to test whether the sky coverage of {\sl \textsl{APOGEE}} could bias the measurement results. Analysis of this mock reveals that our method is largely footprint-agnostic in the parameters of interest. We recover the same components in the mock without the footprint constraints, including each angular momentum parameter being within 3$\sigma$ of the corresponding mock with the footprint constraint.

In summary, the mock {\sl \textsl{APOGEE}} data set provides an appropriately similar testing ground to the {\sl \textsl{APOGEE}} data, demonstrating that our methodology for dynamically constraining different components is robust. The mock, including both with and without the \textsl{\textsl{APOGEE}} sightlines, is available at \url{https://github.com/michael-petersen/BarGMM/tree/main}.

\section{Gaussian mixture-model fit results}
\label{appen_two}

In Table~\ref{tab:fitparams}, we list the Cartesian parameters for the full output of the GMM decompositions. For completeness, we list all fit components, even when statistically insignificant ($f\lesssim0.01$). For the statistically insignificant components, we do not assign an interpretation.

\begin{table*}
\centering
  \begin{tabular}{l|l|l|l|l|l|l|l|l|l}
  \hline
 name & radii& $f$ & L$_x$ & L$_y$ & L$_z$ & $\gamma$ & $\sigma_x$ & $\sigma_y$ & $\sigma_z$ \\
 &[kpc] & [$\%$] & [$\kpc~\kms$]& [$\kpc~\kms$] & [$\kpc~\kms$] & [$^\circ$] & [$\kpc~\kms$] & [$\kpc~\kms$] &  [$\kpc~\kms$] \\
\hline
disc&$0.0-1.0$&$0.1596^{+0.0262}_{-0.0258}$&$1.8^{+12.4}_{-12.7}$&$9.3^{+12.0}_{-12.3}$&$92.8^{+10.1}_{-9.5}$&$24.2^{+16.5}_{-54.0}$&$130.4^{+14.1}_{-10.4}$&$142.3^{+13.7}_{-14.4}$&$41.7^{+6.3}_{-6.2}$\\
bar&$0.0-1.0$&$0.6136^{+0.0286}_{-0.0274}$&$2.5^{+3.2}_{-3.1}$&$2.2^{+2.4}_{-2.5}$&$11.7^{+2.8}_{-2.8}$&$38.8^{+3.6}_{-3.6}$&$80.3^{+3.0}_{-2.8}$&$55.7^{+3.0}_{-2.8}$&$30.8^{+2.5}_{-2.2}$\\
knot&$0.0-1.0$&$0.2274^{+0.0207}_{-0.0203}$&$-0.7^{+2.3}_{-1.0}$&$-0.8^{+1.5}_{-0.9}$&$0.0^{+0.8}_{-0.8}$&$33.5^{+5.1}_{-4.3}$&$25.8^{+2.8}_{-2.6}$&$10.5^{+0.9}_{-0.4}$&$10.1^{+0.2}_{-0.1}$\\
    \hline
disc&$0.5-1.5$&$0.1872^{+0.0207}_{-0.0212}$&$-3.7^{+10.0}_{-10.1}$&$9.3^{+10.7}_{-10.4}$&$117.1^{+7.9}_{-7.7}$&$31.6^{+12.4}_{-13.3}$&$150.3^{+8.4}_{-7.3}$&$173.8^{+10.4}_{-9.2}$&$69.4^{+6.3}_{-6.0}$\\
bar&$0.5-1.5$&$0.6185^{+0.0224}_{-0.0213}$&$2.5^{+3.0}_{-2.9}$&$2.8^{+2.4}_{-2.4}$&$25.5^{+2.9}_{-2.8}$&$35.4^{+3.6}_{-3.7}$&$83.2^{+2.7}_{-2.6}$&$59.9^{+2.6}_{-2.5}$&$42.0^{+2.2}_{-2.1}$\\
knot&$0.5-1.5$&$0.1941^{+0.0152}_{-0.0156}$&$-0.7^{+1.9}_{-1.5}$&$-3.0^{+1.4}_{-1.4}$&$0.3^{+0.9}_{-0.9}$&$31.8^{+4.6}_{-4.2}$&$25.2^{+2.3}_{-2.2}$&$10.6^{+1.0}_{-0.5}$&$10.1^{+0.2}_{-0.1}$\\
    \hline
disc&$1.0-2.0$&$0.3802^{+0.023}_{-0.0225}$&$-7.9^{+6.6}_{-6.4}$&$-1.2^{+7.7}_{-7.5}$&$160.8^{+6.4}_{-6.3}$&$44.7^{+29.0}_{-29.4}$&$154.2^{+5.3}_{-5.2}$&$158.6^{+5.7}_{-5.3}$&$89.5^{+4.4}_{-4.2}$\\
bar&$1.0-2.0$&$0.5005^{+0.0222}_{-0.0221}$&$2.9^{+3.2}_{-3.4}$&$-4.0^{+2.5}_{-2.6}$&$-0.6^{+3.5}_{-3.1}$&$40.1^{+3.9}_{-4.0}$&$80.5^{+2.9}_{-2.7}$&$56.3^{+2.9}_{-2.8}$&$44.2^{+2.6}_{-2.5}$\\
knot&$1.0-2.0$&$0.1198^{+0.0126}_{-0.0119}$&$0.1^{+0.8}_{-1.2}$&$-0.4^{+1.5}_{-0.6}$&$0.0^{+0.7}_{-0.8}$&$16.5^{+9.2}_{-9.4}$&$18.5^{+2.4}_{-2.0}$&$10.6^{+0.9}_{-0.5}$&$10.3^{+0.5}_{-0.2}$\\
    \hline
disc&$1.5-2.5$&$0.5688^{+0.0367}_{-0.0312}$&$-7.6^{+5.6}_{-5.6}$&$-4.3^{+5.3}_{-5.2}$&$210.8^{+7.0}_{-6.1}$&$52.8^{+12.9}_{-12.2}$&$168.0^{+4.7}_{-4.4}$&$155.4^{+4.9}_{-4.3}$&$128.4^{+4.6}_{-4.2}$\\
bar&$1.5-2.5$&$0.3506^{+0.0293}_{-0.0325}$&$4.0^{+4.6}_{-4.7}$&$-6.2^{+3.4}_{-3.4}$&$-9.6^{+8.0}_{-6.9}$&$58.5^{+4.3}_{-4.1}$&$89.3^{+5.0}_{-4.8}$&$58.0^{+4.9}_{-4.5}$&$59.9^{+5.5}_{-5.2}$\\
knot&$1.5-2.5$&$0.0818^{+0.01}_{-0.0095}$&$0.3^{+0.8}_{-1.8}$&$0.4^{+0.7}_{-1.5}$&$0.0^{+1.0}_{-1.1}$&$31.0^{+8.7}_{-7.8}$&$20.7^{+2.9}_{-2.7}$&$10.4^{+0.8}_{-0.3}$&$10.9^{+1.4}_{-0.7}$\\
    \hline
disc&$2.0-3.0$&$0.4845^{+0.034}_{-0.0366}$&$-10.6^{+6.6}_{-6.6}$&$-3.1^{+7.4}_{-6.5}$&$369.5^{+10.3}_{-10.4}$&$67.1^{+32.7}_{-14.2}$&$187.6^{+5.3}_{-5.4}$&$178.5^{+6.3}_{-5.6}$&$140.6^{+7.4}_{-6.5}$\\
bar&$2.0-3.0$&$0.4708^{+0.0357}_{-0.0332}$&$6.2^{+5.5}_{-5.3}$&$-1.6^{+3.5}_{-3.6}$&$156.7^{+8.7}_{-9.6}$&$75.8^{+2.4}_{-2.5}$&$140.3^{+4.8}_{-4.9}$&$88.0^{+3.8}_{-4.1}$&$136.8^{+4.6}_{-5.1}$\\
knot&$2.0-3.0$&$0.0443^{+0.0066}_{-0.0074}$&$2.0^{+3.5}_{-4.0}$&$-3.6^{+2.1}_{-2.1}$&$6.6^{+3.9}_{-3.7}$&$58.7^{+8.2}_{-7.0}$&$28.0^{+7.3}_{-6.1}$&$11.2^{+1.9}_{-0.9}$&$19.6^{+4.7}_{-3.7}$\\
    \hline
-&$2.5-3.5$&$0.0113^{+0.002}_{-0.0024}$&$-16.8^{+73.8}_{-61.6}$&$-365.2^{+35.2}_{-46.7}$&$-316.4^{+34.9}_{-36.8}$&$21.2^{+11.7}_{-13.2}$&$238.0^{+37.7}_{-28.6}$&$387.6^{+57.0}_{-53.6}$&$142.4^{+28.1}_{-22.4}$\\
disc&$2.5-3.5$&$0.6113^{+0.0325}_{-0.035}$&$-14.3^{+5.0}_{-5.0}$&$-4.3^{+3.7}_{-2.9}$&$476.9^{+9.1}_{-8.9}$&$78.6^{+5.5}_{-4.9}$&$195.8^{+3.6}_{-3.9}$&$171.9^{+4.2}_{-3.7}$&$158.0^{+4.7}_{-4.6}$\\
bar&$2.5-3.5$&$0.3774^{+0.0351}_{-0.0326}$&$2.0^{+5.8}_{-5.5}$&$0.3^{+3.3}_{-3.0}$&$215.3^{+11.8}_{-12.2}$&$80.7^{+1.6}_{-1.6}$&$157.9^{+4.7}_{-4.8}$&$77.5^{+3.8}_{-4.3}$&$148.6^{+6.1}_{-5.8}$\\
    \hline
disc&$3.0-4.0$&$0.6373^{+0.0302}_{-0.0305}$&$-20.4^{+4.3}_{-4.5}$&$-2.0^{+2.4}_{-1.4}$&$616.3^{+8.1}_{-8.3}$&$84.7^{+4.6}_{-3.6}$&$197.4^{+3.2}_{-3.1}$&$177.2^{+3.6}_{-3.2}$&$168.2^{+4.2}_{-3.9}$\\
-&$3.0-4.0$&$0.0157^{+0.0022}_{-0.0024}$&$32.7^{+36.9}_{-35.6}$&$-197.3^{+36.0}_{-39.8}$&$-257.5^{+27.8}_{-25.4}$&$22.9^{+7.5}_{-7.3}$&$206.7^{+22.6}_{-20.5}$&$355.5^{+36.3}_{-33.9}$&$165.9^{+22.7}_{-20.1}$\\
bar&$3.0-4.0$&$0.347^{+0.0313}_{-0.0303}$&$-9.1^{+6.1}_{-6.0}$&$-2.3^{+3.3}_{-3.2}$&$318.3^{+13.4}_{-14.0}$&$78.9^{+1.3}_{-1.3}$&$186.7^{+5.1}_{-4.8}$&$87.7^{+3.6}_{-3.9}$&$174.3^{+6.9}_{-6.4}$\\
    \hline
-&$3.5-4.5$&$0.0003^{+0.0002}_{-0.0002}$&$94.9^{+286.7}_{-260.9}$&$-528.0^{+54.6}_{-91.1}$&$-924.7^{+74.0}_{-75.3}$&$43.1^{+29.3}_{-29.4}$&$13.7^{+8.4}_{-2.8}$&$14.0^{+9.1}_{-3.1}$&$14.1^{+9.7}_{-3.1}$\\
-&$3.5-4.5$&$0.0003^{+0.0002}_{-0.0002}$&$-315.3^{+246.7}_{-350.6}$&$549.0^{+241.1}_{-94.8}$&$-375.8^{+38.5}_{-39.1}$&$44.1^{+28.7}_{-30.8}$&$14.5^{+10.3}_{-3.5}$&$14.3^{+9.3}_{-3.3}$&$14.2^{+9.3}_{-3.2}$\\
disc&$3.5-4.5$&$0.9994^{+0.0003}_{-0.0002}$&$-16.7^{+2.6}_{-2.8}$&$5.3^{+2.0}_{-2.1}$&$638.5^{+3.6}_{-3.6}$&$82.6^{+1.6}_{-1.6}$&$204.2^{+1.7}_{-1.9}$&$160.9^{+1.6}_{-1.6}$&$258.7^{+2.5}_{-2.4}$\\
    \hline
-&$4.0-5.0$&$0.0002^{+0.0001}_{-0.0002}$&$-443.2^{+380.9}_{-803.6}$&$-612.7^{+298.2}_{-380.7}$&$-406.0^{+373.3}_{-365.4}$&$35.8^{+14.5}_{-17.1}$&$10.4^{+0.3}_{-0.3}$&$10.5^{+0.6}_{-0.3}$&$11.4^{+0.4}_{-0.4}$\\
-&$4.0-5.0$&$0.0004^{+0.0002}_{-0.0003}$&$355.9^{+121.5}_{-102.1}$&$638.2^{+84.6}_{-108.2}$&$-395.1^{+258.5}_{-27.2}$&$9.4^{+5.8}_{-8.6}$&$17.3^{+8.5}_{-3.7}$&$11.0^{+1.1}_{-0.7}$&$24.2^{+14.9}_{-6.7}$\\
disc&$4.0-5.0$&$0.9994^{+0.0004}_{-0.0002}$&$-7.6^{+1.5}_{-1.6}$&$10.9^{+0.8}_{-0.8}$&$792.7^{+2.8}_{-2.8}$&$83.9^{+0.8}_{-0.6}$&$205.6^{+1.3}_{-1.2}$&$152.4^{+0.6}_{-0.6}$&$280.8^{+1.0}_{-0.9}$\\
  \hline
  \hline

\end{tabular}
  \caption{Values for parameters of interest derived from the GMM fits.}
  \label{tab:fitparams}
\end{table*}

\bsp	
\label{lastpage}
\end{document}

%% file: preamble.tex
\usepackage{savesym}
\savesymbol{tablenum}
\usepackage{siunitx}
\restoresymbol{SIX}{tablenum}
\DeclareSIUnit\year{yr}
\DeclareSIUnit\parsec{pc}
\DeclareSIUnit\msun{M_\odot}
\DeclareSIUnit\Rsun{R_\odot}

\newcommand{\kms}{\unit{\km\per\s}}
\newcommand{\kpc}{\unit{\kilo\parsec}}
\newcommand{\pc}{\unit{\parsec}}

\DeclareSIUnit\mstar{M_\star}

\newcommand{\logg}{\ensuremath{\log g}}

\usepackage{soul}
\newcounter{MPcounter}

\newcommand{\boldmatrix}[1]{\boldsymbol{\mathsf{#1}}}
